\documentstyle[11pt,aaspp4]{article}
\begin{document}

\title{DIRECT Distances to Nearby Galaxies Using Detached Eclipsing 
Binaries and Cepheids. I. Variables in the Field M31B\footnote{Based
on the observations collected at the Michigan-Dartmouth-MIT (MDM)
1.3-meter telescope and at the F. L. Whipple Observatory (FLWO)
1.2-meter telescope}}

\author{J. Kaluzny}
\affil{Warsaw University Observatory, Al. Ujazdowskie 4,
00--478 Warszawa, Poland} 
\affil{\tt e-mail: jka@sirius.astrouw.edu.pl} 
\author{K. Z. Stanek\altaffilmark{2}, M. Krockenberger, D. D. Sasselov} 
\affil{\tt e-mail: kstanek@cfa.harvard.edu, krocken@cfa.harvard.edu,
sasselov@cfa.harvard.edu} 
\affil{Harvard-Smithsonian Center for Astrophysics, 60 Garden St., 
Cambridge, MA~02138} 
\altaffiltext{2}{On leave from N.~Copernicus Astronomical Center, 
Bartycka 18, Warszawa 00--716, Poland} 
\author{J. L. Tonry} 
\affil{University of Hawaii, Institute for Astronomy, 
2680 Woodlawn Dr., Honolulu, HI~96822}
\affil{\tt e-mail: jt@avidya.ifa.hawaii.edu} 
\author{M. Mateo}
\affil{Department of Astronomy, University of Michigan, 
821 Dennison Bldg., Ann Arbor, MI~48109--1090} 
\affil{\tt e-mail: mateo@astro.lsa.umich.edu}

\begin{abstract}

We undertook a long term project, DIRECT, to obtain the direct
distances to two important galaxies in the cosmological distance
ladder -- M31 and M33, using detached eclipsing binaries (DEBs) and
Cepheids. While rare and difficult to detect, detached eclipsing
binaries provide us with the potential to determine these distances
with an accuracy better than 5\%. The
massive photometry obtained in order to detect DEBs provides us with
good light curves for the Cepheid variables. These are essential to
the parallel project to derive direct Baade-Wesselink distances to
Cepheids in M31 and M33. For both Cepheids and eclipsing binaries the
distance estimates will be free of any intermediate steps.

As a first step of the DIRECT project, between September 1996 and
January 1997 we have obtained 36 full nights on the
Michigan-Dartmouth-MIT (MDM) 1.3-meter telescope and 45 full/partial
nights on the F. L. Whipple Observatory (FLWO) 1.2-meter telescope to
search for detached eclipsing binaries and new Cepheids in the M31 and
the M33 galaxies.  In this paper, first in the series, we present the
catalog of variable stars, most of them newly detected, found in the
field M31B ($\alpha_{2000.0},\delta_{2000}=11.20\deg,41.59\deg$).
We have found 85 variable stars: 12 eclipsing binaries, 38 Cepheids
and 35 other periodic, possible long period or non-periodic
variables. The catalog of variables, as well as their photometry and
finding charts, are available using the {\tt anonymous ftp} service
and the {\tt WWW}.

\end{abstract}


\section{Introduction}

The two nearby galaxies -- M31 and M33, are stepping stones to most of
our current effort to understand the evolving universe at large
scales.  First, they are essential to the calibration of the
extragalactic distance scale (Jacoby et al.~1992; Tonry et
al.~1997). Second, they constrain population synthesis models for
early galaxy formation and evolution, and provide the stellar
luminosity calibration. There is one simple requirement for all this
-- accurate distances.
 
Detached eclipsing binaries (DEBs) have the potential to establish
distances to M31 and M33 with an unprecedented accuracy of better than
5\% and possibly to better than 1\%. These distances are now known to
no better than 10-15\%, as there are discrepancies of $0.2-0.3\;{\rm
mag}$ between RR Lyrae and Cepheids distance indicators (e.g.~Huterer,
Sasselov \& Schechter 1995).  Detached eclipsing binaries (for reviews
see Andersen 1991, Paczy\'nski 1997) offer a single step distance
determination to nearby galaxies and may therefore provide an accurate
zero point calibration -- a major step towards very accurate
determination of the Hubble constant, presently an important but
daunting problem for astrophysicists (see the papers from the recent
``Debate on the Scale of the Universe'': Tammann 1996, van den Bergh
1996).
 
The detached eclipsing binaries have yet to be used (Huterer, Sasselov
\& Schechter 1995; Hilditch 1996) as distance indicators to M31 and
M33. According to Hilditch (1996), there are about 60 eclipsing
binaries of all kinds known in M31 (Gaposchkin 1962; Baade \& Swope
1963; Baade \& Swope 1965) and only {\em one} in M33 (Hubble 1929)!
Only now does the availability of large format CCD detectors and
inexpensive CPUs make it possible to organize a massive search for
periodic variables, which will produce a handful of good DEB
candidates. These can then be spectroscopically followed-up with the
powerful new 6.5-10 meter telescopes.

The study of Cepheids in M31 and M33 has a venerable history (Hubble
1926, 1929; Gaposchkin 1962; Baade \& Swope 1963; Baade \& Swope
1965). In the 80's Freedman \& Madore (1990) and Freedman, Wilson \&
Madore (1991) studied small samples of the earlier discovered
Cepheids, to build PL relations in M31 and M33, respectively.
However, both the sparse photometry and the small samples do not
provide a good basis for obtaining direct BW distances to Cepheids --
the need for new digital photometry has been long overdue. Recently,
Magnier et al.~(1997) surveyed large portions of M31, which have
previously been ignored, and found some 130 new Cepheid variable
candidates.  Their light curves are however rather sparsely sampled
and in $V$ band only.

In this paper, first of the series, we present the catalog of variable
stars, most of them newly detected, found in one of the fields in
M31. In Sec.2 we discuss the selection of the fields in M31 and the
observations. In Sec.3 we describe the data reduction and
calibration. In Sec.4 we discuss the automatic selection we used for
finding the variable stars. In Sec.5 we discuss the classification of
the variables, also using well-defined algorithms whenever possible.
In Sec.6 we present the catalog of variable stars. Finally, in Sec.7
we discuss the future follow-up observations and research necessary to
fully explore the potential offered by DEBs and Cepheids as direct
distance indicators.

\section{Fields selection and observations}

M31 was primarily observed with the McGraw-Hill 1.3-meter telescope at
the MDM Observatory. We used the front-illuminated, Loral $2048^2$ CCD
Wilbur (Metzger, Tonry \& Luppino 1993), which at the $f/7.5$ station
of the 1.3-meter has a pixel scale of $0.32\;arcsec/pixel$ and field
of view of roughly $11\;arcmin$. We used Kitt Peak Johnson-Cousins
$BVI$ filters.  Some data for M31 were also obtained with the
1.2-meter telescope at the FLWO, where we used ``AndyCam'' with
thinned, back-side illuminated, AR coated Loral $2048^2$ CCD.  The
pixel scale happens to be essentially the same as at the MDM 1.3-meter
telescope. We used standard Johnson-Cousins $BVI$ filters.

\begin{figure}[t]
\plotfiddle{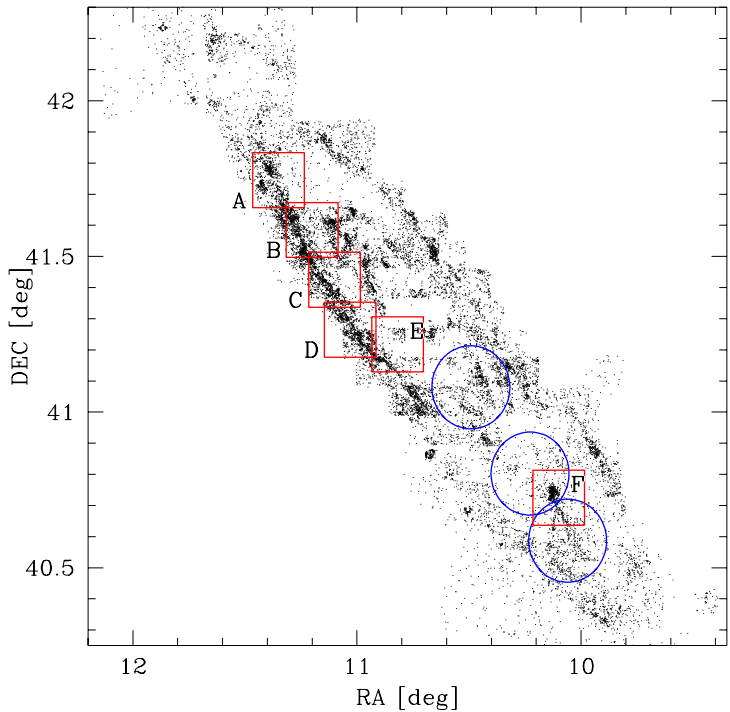}{8cm}{0}{50}{50}{-160}{-85}
\caption{Fields observed in M31 during Fall 1996
(squares), based on the photometric survey of M31 by Magnier et
al.~(1992) and Haiman et al.~(1994). Only blue stars, with $B-V<0.4$,
are shown.  Also shown (circles) are Baade's Fields I, II, III.}
\label{fig:fields}
\end{figure}

Fields in M31 were selected using the MIT photometric survey of M31 by
Magnier et al.~(1992) and Haiman et al.~(1994). Fig.\ref{fig:fields}
shows stars from this survey with $B-V<0.4$, i.e.~blue stars belonging
to M31. We selected six $11'\times11'$ fields, M31A--F, four of them
(A--D) concentrated on the rich spiral arm, one (E) coinciding with the
region of M31 searched for microlensing by Crotts \& Tomaney (1996),
and one (F) containing the giant star formation region known as NGC206
(observed by Baade \& Swope 1963). Fields A--C were observed during
September and October 1996 5--8 times per night in the $V$ band,
resulting in total of 130-160 V exposures per field. Fields D--F were
observed once a night in the $V$ band. Some exposures in $B$ and $I$
bands were also taken. M31 was also occasionally observed at the FLWO
1.2-meter telescope, whose main target was M33.

In this paper we present the results for the most frequently observed
field, M31B.  We obtained for this field useful data during 29 nights
at the MDM, collecting a total of 160 $900\;sec$ exposures in $V$, 27
$600\;sec$ exposures in $I$ and 2 $1200\;sec$ exposures in $B$. We
also obtained for this field useful data during 14 nights at the FLWO,
collecting a total of 4 $900\;sec$ exposures in $V$ and 17 $600\;sec$
exposures in $I$. The complete list of exposures for this field
and related data files are available through {\tt anonymous
ftp} on {\tt cfa-ftp.harvard.edu}, in {\tt pub/kstanek/DIRECT}
directory. Please retrieve the {\tt README} file for instructions.
Additional information on the DIRECT project is available through the
{\tt WWW} at {\tt http://cfa-www.harvard.edu/\~\/kstanek/DIRECT/}.

\section{Data reduction, calibration and astrometry}

\subsection{Initial reduction, PSF fitting}

Preliminary processing of the CCD frames was done with the standard
routines in the IRAF-CCDPROC package.\footnote{IRAF is distributed by
the National Optical Astronomy Observatories, which are operated by
the Associations of Universities for Research in Astronomy, Inc.,
under cooperative agreement with the NSF} For all filters the
flatfield frames were prepared by combining ``dome flats" and
exposures of the twilight sky. These reductions reduced total
instrumental systematics to below 1\%.  The bad columns and hot/cold
pixels were masked out using the IRAF routine IMREPLACE.

Stellar profile photometry was extracted using the {\it
Daophot/Allstar} package (Stetson 1987, 1991).  The analyzed images
showed a significant positional dependence of the point spread
function (PSF), which was well fit by a Moffat-function PSF,
quadratically varying with $X,Y$.  We selected a ``template'' frame
for each filter using a single frame of particularly good quality.
These template images were reduced in a standard way. A set of
approximately 100 relatively isolated stars was selected to build the
PSF for each image.  The PSF star lists as well as lists of objects
measured on template images were then used for reduction of remaining,
``non-template'', images. For each individual image we first ran FIND
and PHOT programs to obtain a preliminary list of stellar positions,
then the stars from the ``master'' PSF list for a given filter were
automatically identified, and the PSF was derived. Next for each frame
we executed the {\it Allstar} program to obtain improved positions for
the stars.  These positions were used to transform coordinates of the
stars included on the ``master'' list into the coordinates of the
current frame. {\it Allstar} was then ran in the fixed-position-mode
using as an input the transformed ``master'' list, and the resulting
output file contained photometry only for stars measured on the
``template" images. There are two classes of objects which may be
missed: a) objects located outside "template'' images but inside the
present image; and b) objects located inside the ``template'' field
but not included on the master list.  By carefully positioning the
telescope the offsets between images were small, and in most cases did
not exceed 15 pixels.  We were, however, concerned about potential
variables, such as novae, which could be un-measurable on ``template''
frames but measurable on some fraction of images. To avoid losing such
objects we updated the master list by adding object found by {\it
Daophot/Allstar} in the ``non-fixed-position'' mode, detected above
$10\sigma$ threshold in the residual images left after subtracting the
objects on the current ``master'' list. Next, {\it Allstar} was ran
again in the ``non-fixed-position'' mode using the extended list of
stars.  Some additional fraction of faint ``template'' objects was
usually rejected by {\it Allstar} at this step.  As the end result of
this procedure we had for each of processed frames (with exception of
template images) two lists of photometry: one list including
exclusively ``template'' objects and one including mixture of
``template'' and ``non-template'' stars.

Both lists of instrumental photometry derived for a given frame were
transformed to the common instrumental system of the appropriate
``template'' image. As it turned out the offsets of instrumental
magnitudes were slightly position dependent and changed by
$<0.04\;{\rm mag}$ across a field. This effect was taken into account
while transforming photometry to the instrumental system of
``template'' images. We traced down the source of this problem and
found that it is caused by non-perfect modeling of variable PSF in the
corners of the images. This means also that photometry derived from
``template'' images is affected by systematic, position dependent
errors. The problem could be cured if it were possible to determine
accurate aperture corrections for a large number of stars distributed
uniformly over the whole field. Unfortunately this was not the case
with our images of M31. The observed fields are very crowded and their
images contain limited number of stars with very high S/N.  To
estimate the size of systematic errors in our photometry we analyzed a
set of images of the open cluster NGC~6791 also taken with the MDM
1.3-meter telescope (Kaluzny et al.~1997). The images of NGC~6791 are
moderately crowded and contain few hundred stars with S/N sufficiently
high for the determination of aperture corrections. Based on
examination of the derived aperture corrections and on comparison of
profile photometry of NGC~6791 with the photometry of this cluster
available in literature (Kaluzny \& Rucinski 1995) we concluded that
the systematic errors for stars located in corners of M31 fields are
$<0.05\;{\rm mag}$.

Photometry obtained for $V$ and $I$ filters was combined into data
bases. Two data bases were prepared for each of the filters. One
included only photometry for the ``template'' stars obtained by
running {\it Allstar} in a ``fixed-position-mode", and second included
mixture of ``template'' and ``non-template'' objects and was obtained
by running {\it Allstar} in the ``non-fixed-position'' mode. In this
paper we search for variables only in the first database, i.e.~for the
``template'' stars only.

\subsection{Photometric calibration and astrometry}

\begin{figure}[p]
\plotfiddle{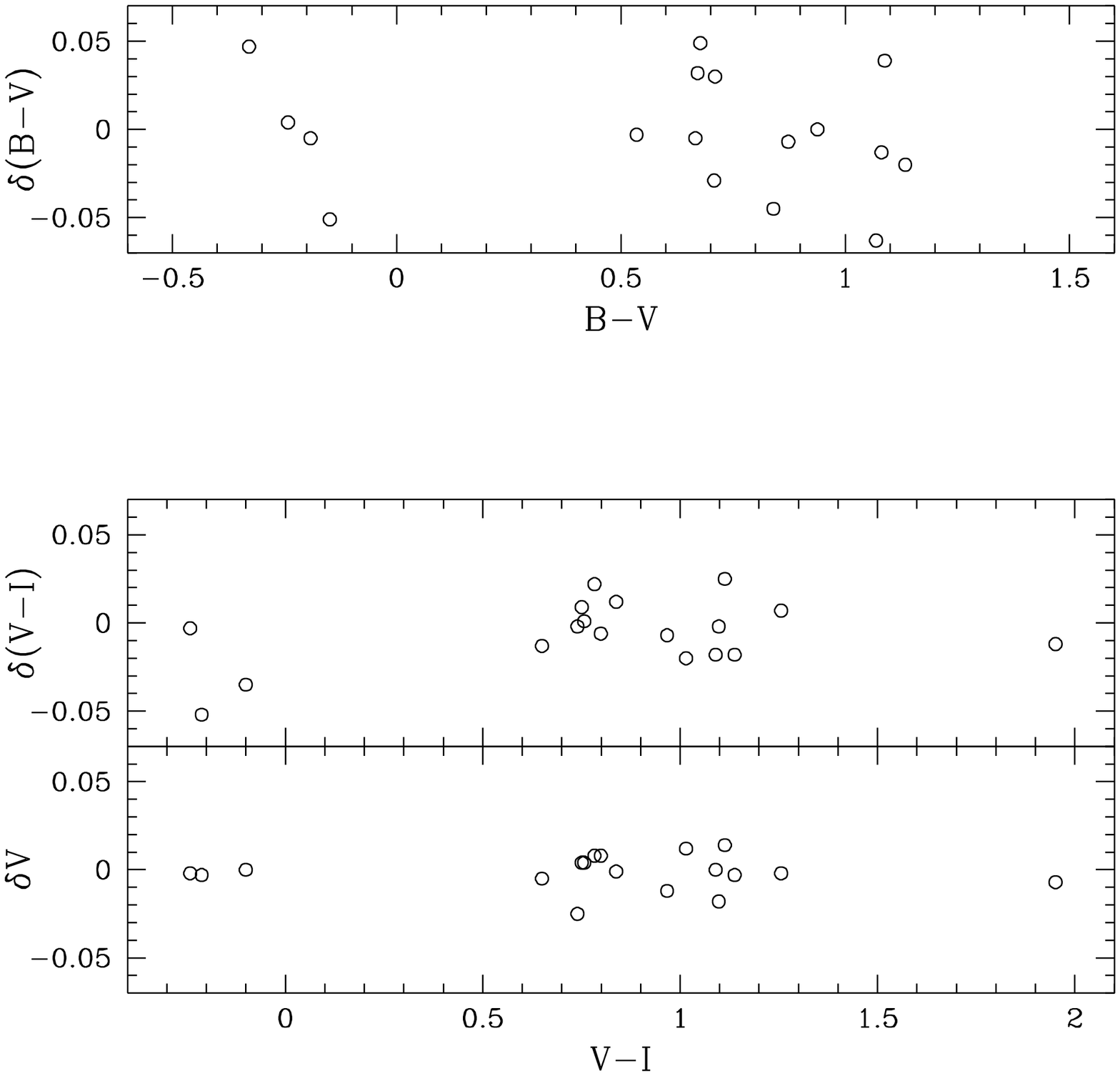}{8cm}{0}{50}{50}{-160}{-85}
\caption{Residuals between the standard and calculated magnitudes and 
colors for 18 standard stars (Landolt 1992) observed on Sept.~14/15,
1996.}
\label{fig:landolt}
\plotfiddle{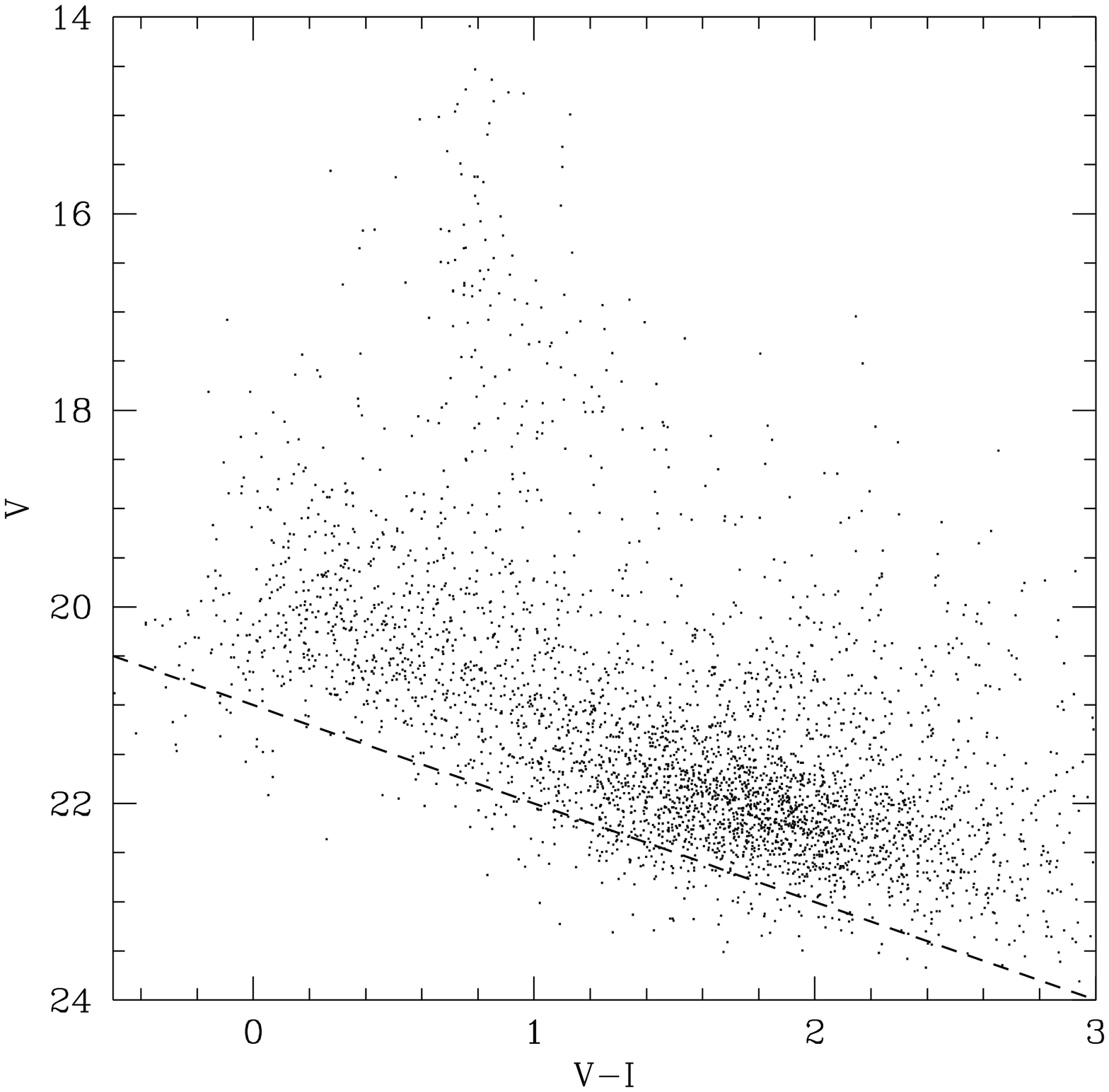}{9cm}{0}{50}{50}{-160}{-85}
\caption{$V, V-I$ color-magnitude diagram for $\sim3,700$
stars in the field M31B. The dashed line corresponds to the $I$
detection limit of $I\sim21\;{\rm mag}$. }
\label{fig:cmd}
\end{figure}

On the night of Sept.~14/15, 1996 we observed 4 Landolt (1992) fields
containing a total of 18 standards stars. These fields were observed
through the $BV$ filters at air-masses ranging from 1.12 to 1.75, and
through the $I$ filter at air masses ranging from 1.12 to 1.53.  The
following transformation from the instrumental to the standard system
was derived:
\begin{eqnarray}
b-v=0.231+0.648(B-V)+0.15X\\
v=V+3.128-0.007(V-I)+0.13X\\
v-i=0.152+1.007(V-I)+0.08X\\
i=I+2.969-0.010(V-I)+0.05X
\end{eqnarray}
where lower case letters correspond to the instrumental magnitudes and
$X$ is the air mass. It was possible to derive with confidence
extinction coefficient for the $V$ filter only. Extinction
coefficients for the $B$ and $I$ filters were assumed.  In
Fig.\ref{fig:landolt} we show the residuals between the standard and
calculated magnitudes and colors for the standard stars. The derived
transformation satisfactorily reproduces the $V$ and $I$ magnitudes
and $V-I$ colors. The $B-V$ transformation reproduces the standard
system poorly, due to a rapid decline of quantum efficiency of the
Wilbur CCD camera in the range of wavelengths corresponding to the $B$
band. We therefore decided to drop the $B$ data from our analysis of
M31B, especially since we took only 2 $B$ frames for this field.  We
note that all frames of M31 used for calibration of $VI$ photometry
were obtained in parallel with observations of Landolt standards and
over the air masses not exceeding 1.25. Hence, the fact that we used
assumed extinction for the $I$ band is unlikely to introduce any error
of the zero point exceeding $0.01\;{\rm mag}$ into M31 photometry. In
fact the dominant error of the zero points of the $VI$ photometry for
M31 fields are uncertainties of aperture corrections and systematic
errors of profile photometry for stars positioned in the corners of
the images. We estimate that these external errors of $V$ and $I$
magnitudes are not worse than $0.05\;{\rm mag}$.

The $V,\;V-I$ color-magnitude diagram based on photometry extracted from
the ``template'' images is shown in Fig.\ref{fig:cmd}. The dashed line
corresponds to the $I$ detection limit of $I\sim21\;{\rm mag}$ (see
the next section). Stars near $V\sim22$ and $V-I\sim1.8$ represent the
top of the evolved red giant population. The vertical strip of stars
between $0.6<V-I<1.2$ and $V<20$ are the Galactic foreground stars.
Stars bluer than $V-I<0.6$ are the upper main sequence, OB type stars,
in M31.

\begin{figure}[t]
\plotfiddle{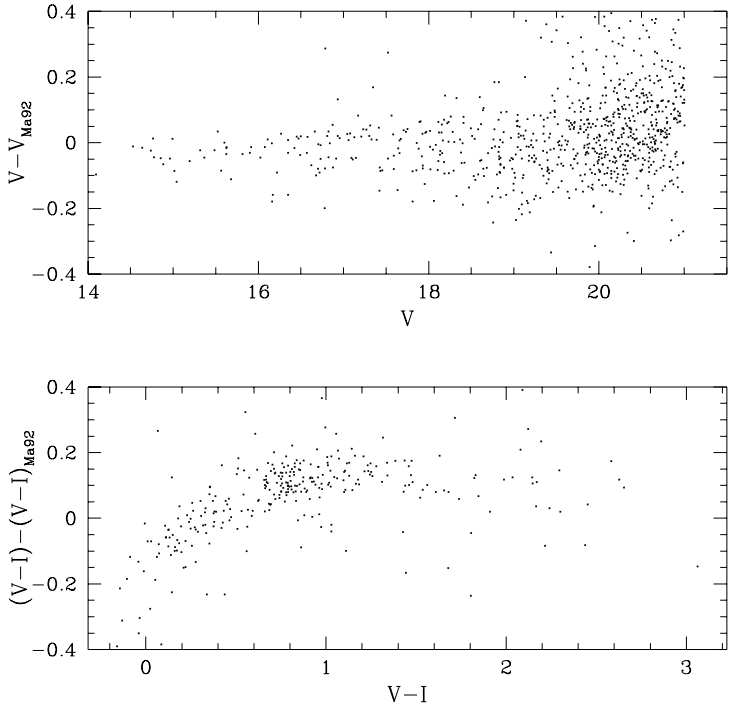}{8cm}{0}{50}{50}{-160}{-85}
\caption{Residuals between the standard $V$ and $V-I$
obtained for the same stars by us and by Magnier et al.~(1992).  The
agreement for the $V$ values is very good, while there is a strong
discrepancy in the values of $V-I$. For discussion see text.}
\label{fig:magnier}
\end{figure}

We decided to verify our photometric calibration by matching our stars
to the photometric survey of Magnier et al.~(1992) (hereafter referred
to by Ma92) and comparing their photometry.  Looking at the upper
panel of Fig.\ref{fig:magnier}, we can see that the $V$ band
photometry matches satisfactorily, and for 92 matched stars with
$V<20$ the average difference between ``our'' $V$ and the $V$ values
measured by Ma92 is $0.013\;{\rm mag}$. On the other hand, there is a
strong disagreement between the $V-I$ colors for 303 common stars
(lower panel of Fig.\ref{fig:magnier}). We therefore decided to
recheck our calibrations using a different set of calibration frames.
During one of the photometric nights (Oct.~2/3, 1996) at the MDM
observatory we took a set of calibration frames with the Charlotte
$1024\times1024$ thinned, backside illuminated CCD, which has a pixel
scale of $0.5\;arcsec/pixel$. These calibration frames were reduced in
the same way as described above for the Wilbur chip and the
transformation from the instrumental to the standard system was
derived. Comparing to the transformation for the Wilbur CCD, there was
an offset of $0.04\;{\rm mag}$ in $V$ and $-0.016\;{\rm mag}$ in
$V-I$. This shift of zero points, similar for $V$ and $I$ photometry,
is mainly due to the uncertainties in the aperture corrections, which
we believe are better derived for the Wilbur CCD, which has smaller
pixels.  Apart from this offset, we do not see anything resembling the
strong trend between the $V-I$ residuals and the $V-I$ color, as seen
in the lower panel of Fig.\ref{fig:magnier}.  This discrepancy
certainly deserves further attention.

Additional consistency check comes from comparing our photometry from
the overlapping regions between the fields (Fig.\ref{fig:fields}). We
compared $\sim170$ common stars in the overlap region between the
fields M31B and M31C.  There was an offset of $0.034\;{\rm mag}$ in
$V$ and $0.024\;{\rm mag}$ in $I$, i.e. within our estimate of
$0.05\;{\rm mag}$ discussed above.

Finally, as the last part of the calibration for this field, the
equatorial coordinates were calculated for all objects included in the
data bases for the $V$ filter. The transformation from rectangular
coordinates to equatorial coordinates was derived using 236 stars
identified in the list published by Magnier et al.~(1992), and the
adopted frame solution reproduces equatorial coordinates of these
stars with residuals not exceeding $1.0\;arcsec$.

\section{Selection of variables}

\begin{figure}[t]
\plotfiddle{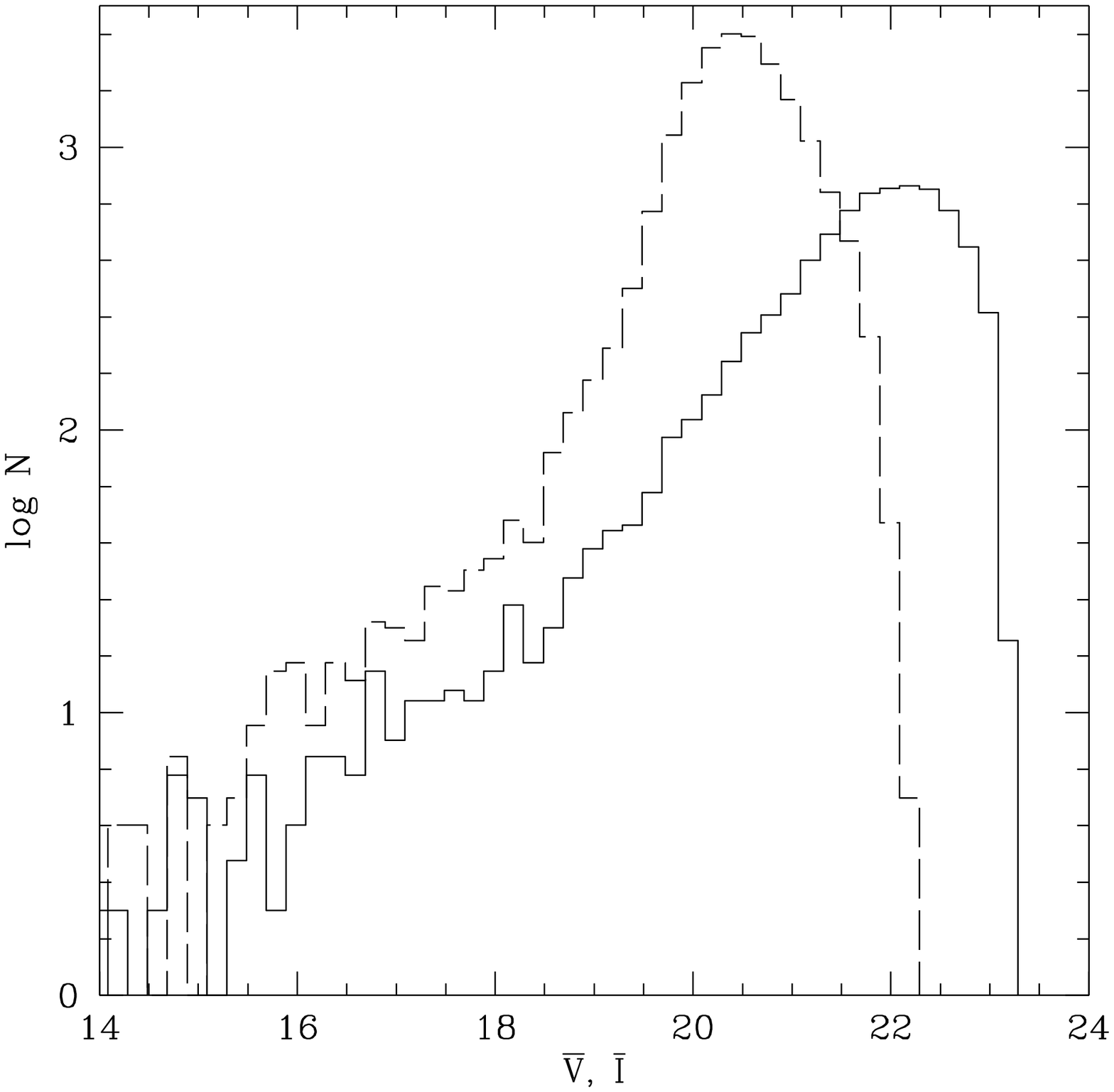}{8cm}{0}{50}{50}{-160}{-85}
\caption{Distributions in $V$ (continuous line) and $I$ (dashed line)
of stars in the field M31B.}
\label{fig:dist}
\end{figure}

The reduction procedure described in Section~3 produces databases of
calibrated $V$ and $I$ magnitudes and their standard errors.  For the
moment we are mostly interested in periodic variable stars, so we use
only the ``first'' database, which includes only ``template'' stars
and was obtained by running {\it Allstar} in a ``fixed-position-mode".
The $V$ database contains 8522 stars with up to 160 measurements, and
the $I$ database contains 18815 stars with up to 27
measurements. Fig.\ref{fig:dist} shows the distributions of stars as a
function of mean $\bar{V}$ and $\bar{I}$ magnitude.  As can be seen
from the shape of the histograms, our completeness starts to drop
rapidly at about $\bar{V}\sim22$ and $\bar{I}\sim20.5$. The primary
reason for this difference in the depth of the photometry between $V$
and $I$ is the level of the combined sky and background light from
unresolved M31 stars, which is about three times higher in the $I$
filter than in the $V$ filter.

\subsection{Removing the ``bad'' points}

The stars measured on each frame are sorted by magnitude, and for each
star we compare its {\em Daophot} errors to those of 300 stars with
similar magnitude located symmetrically on both sides of a given star
in the sorted list. If the {\em Daophot} errors for a given star are
unusually large, the measurement is flagged as ``bad'', and is then
removed when analyzing the lightcurve.  For each star the remaining
measurements are sorted by their error, and the average error and its
standard deviation are calculated. Measurements with errors exceeding
the average error by more than $4\sigma$ are removed, and the whole
procedure is repeated once. Usually 0--10 points are removed, leaving
the majority of stars with roughly $N_{good}\sim150-160$ measurements.
For further analysis we use only those stars which have at least
$N_{good}>N_{max}/2\;(=80)$ measurements. There are 7208 such stars in
the $V$ database of the M31B field.

\subsection{Stetson's variability index}

Our next goal is to select objectively a sample of variable stars from
the total sample defined above.  There are many ways to proceed, and
we will largely follow the approach of Stetson (1996), which is in
turn based on the Welch \& Stetson (1993) algorithm.

We present only a basic summary of Stetson's (1996) procedure (his
Section 2). For each star one can calculate the variability index
\begin{equation}
J=\frac{\sum_{k=1}^{n} w_k sgn(P_k)\sqrt{|{P_k}|}}{\sum_{k=1}^{n} w_k},
\label{eq:stetj}
\end{equation}
where the user has defined $n$ pairs of observations to
be considered, each with a weight $w_k$,
\begin{equation}
P_k=\left\{ \begin{array}{ll}
        \delta_{i(k)}\delta_{j(k)}, 	& \mbox{if $i(k)\neq j(k)$} \\
	\delta_{i(k)}^2 -1, 		& \mbox{if $i(k)=j(k)$} 
\end{array} \right.
\end{equation}
is the product of the normalized residuals of the two paired
observations $i$ and $j$, and
\begin{equation}
\delta=\sqrt{\frac{n}{n-1}}\frac{v-\bar{v}}{\sigma_{v}}
\label{eq:delta}
\end{equation}
is the magnitude residual of a given observation from the average
scaled by the standard error.  There are several nuances in the whole
procedure, and interested reader should consult Stetson's paper for
details.

Following Stetson (1996) we redefine $J$ so it takes into account how
many times a given star was measured.  This is simply done by
multiplying the variability index by a factor $\sum w/w_{max}$, where
$w_{max}$ is the total weight a star would have if measured in all
images.  This gives us the final variability index
\begin{equation}
J_{S}=J\frac{\sum w}{w_{max}}.
\label{eq:stetj1}
\end{equation}
Note that we did not include the measure of the kurtosis of the
magnitude for a given star into the definition of $J_S$, as proposed
by Stetson (1996). We found that including this additional factor made
little change to the total number of stars above certain threshold
$J_{S,min}$, but tended to remove some of the eclipsing variables from
the sample.

To be precise, we should describe how we pair the observations and
what weights $w_k$ we attach to them. Our observing strategy was
designed to have a $V$ image of the M31B field approximately once an
hour, so if two $V$ observations are within $1.5\;hour$ from each
other, we consider them a pair. However, we pair only the subsequent
measurements, so from three closely spaced observations $abc$ we would
select two pairs $ab$ and $bc$, but not $ac$.  In case when $i(k)\neq
j(k)$, we put $w_k=1.0$, in case of $i(k)=j(k)$, we put $w_k=0.25$.
This gives greater weight to longer sequences of closely spaced
observations than to the same number of separated observations, for
example a sequence $abcd$ would have a total weight of 3.0, while a
sequence of $a\;b\;c\;d$ would have the total weight of 1.0.

\subsection{Rescaling of {\em Daophot} errors}

\begin{figure}[t]
\plotfiddle{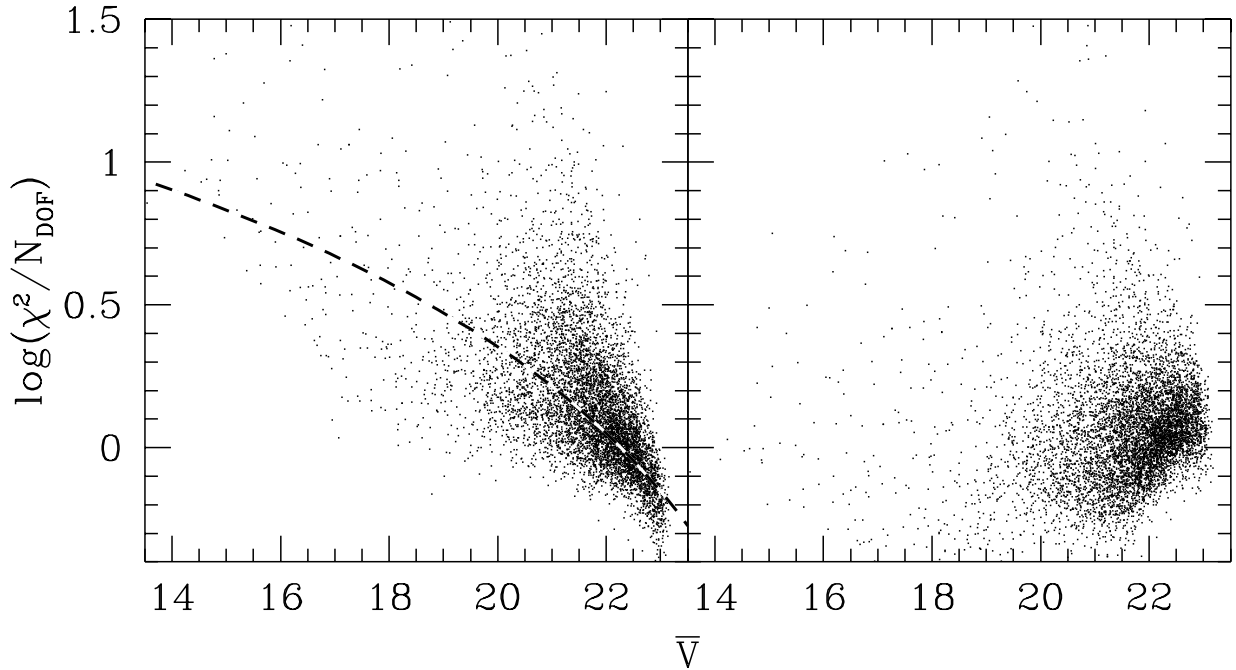}{8cm}{0}{85}{85}{-250}{-340}
\caption{Reduced $\chi^2/N_{DOF}$ vs. average magnitude $\bar{V}$ for
 stars with $N_{good}>80$ measurements. Left panel shows the
uncorrected $\chi^2/N_{DOF}$, the right panel shows the
$\chi^2/N_{DOF}$ after the {\em Daophot} errors were rescaled.  The
average correction to the $\chi^2/N_{DOF}$ is shown in the left panel
with dashed line.  For details see text.}
\label{fig:chi}
\end{figure}

The definition of $\delta$ (Eq.\ref{eq:delta}) includes the standard
errors of individual observations.  If, for some reason, these errors
were over- or underestimated, we would either miss real variables, or
select spurious variables as real ones.  If the standard errors are
over- or underestimated by the same factor, we could easily correct
the results by changing the cutoff value of the variability index
$J_S$ (Eq.\ref{eq:stetj1}). However, this is not the case for our
data.  In the left panel of Fig.\ref{fig:chi} we plot the logarithm of
the $\chi^2/N_{DOF}$ for stars with $N_{good}>80$ measurements.  The
brightest stars ($V\sim15$) have $\chi^2/N_{DOF}\sim10$, so their
errors are underestimated by roughly $\sqrt{10}$, while stars close to
the detection limit, $V\sim23$, have $\chi^2/N_{DOF}<1$ which are too
small.  Whatever the reasons for this correlation, and there are many
possibilities (underestimated flat-fielding errors, less then perfect
PSF fits etc.), we will try to account for the problem in purely
empirical way, by treating the majority of stars as constant, assuming
that for this majority the errors are (roughly) Gaussian.  The
procedure we apply was described in detail by Lupton et al.~(1989),
p.206, and it was used before by Udalski et al.~(1994). We find that
the {\em Daophot} error $\sigma_{D}$ might be used as the real
observational error provided it is multiplied by an appropriate
scaling factor $F$.

\begin{figure}[t]
\plotfiddle{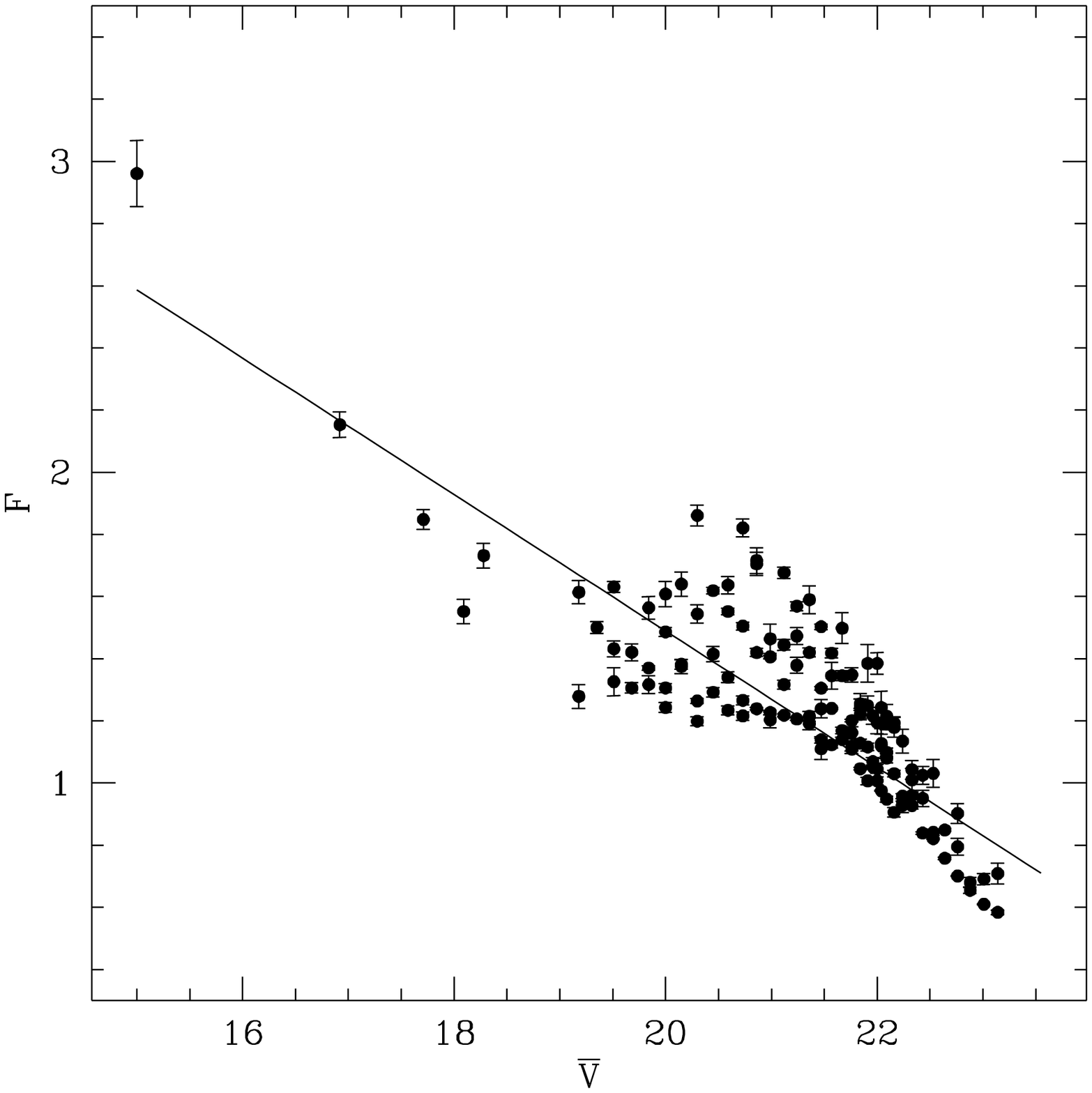}{8cm}{0}{50}{50}{-160}{-85}
\caption{Scaling factor $F$ of {\em Daophot} errors as a function of average 
magnitude $\bar{V}$ (see text for details).}
\label{fig:fsig}
\end{figure}

In Fig.\ref{fig:fsig} we show the scaling factor $F$ as a function of
average magnitude $\bar{V}$. There is a very clear correlation between
the $F$ and $\bar{V}$, to which we fitted a linear relation
$F=5.88-0.22\bar{V}$. In the right panel of Fig.\ref{fig:chi} we plot
the logarithm of the $\chi^2/N_{DOF}$ after the errors were
rescaled. Clearly the distribution now is closer to what one would
expect from Gaussian population with some variable stars
present. However, we do not use rescaled $\chi^2/N_{DOF}$ for
selecting the variable stars. For that we will use the Stetson's $J_S$
(Eq.\ref{eq:stetj1}) instead. Using Stetson's $J_S$ allows to
effectively remove spurious variability caused by few isolated
outstanding points, a property that the $\chi^2$ technique does not
have.

\subsection{Selecting the variables}

\begin{figure}[t]
\plotfiddle{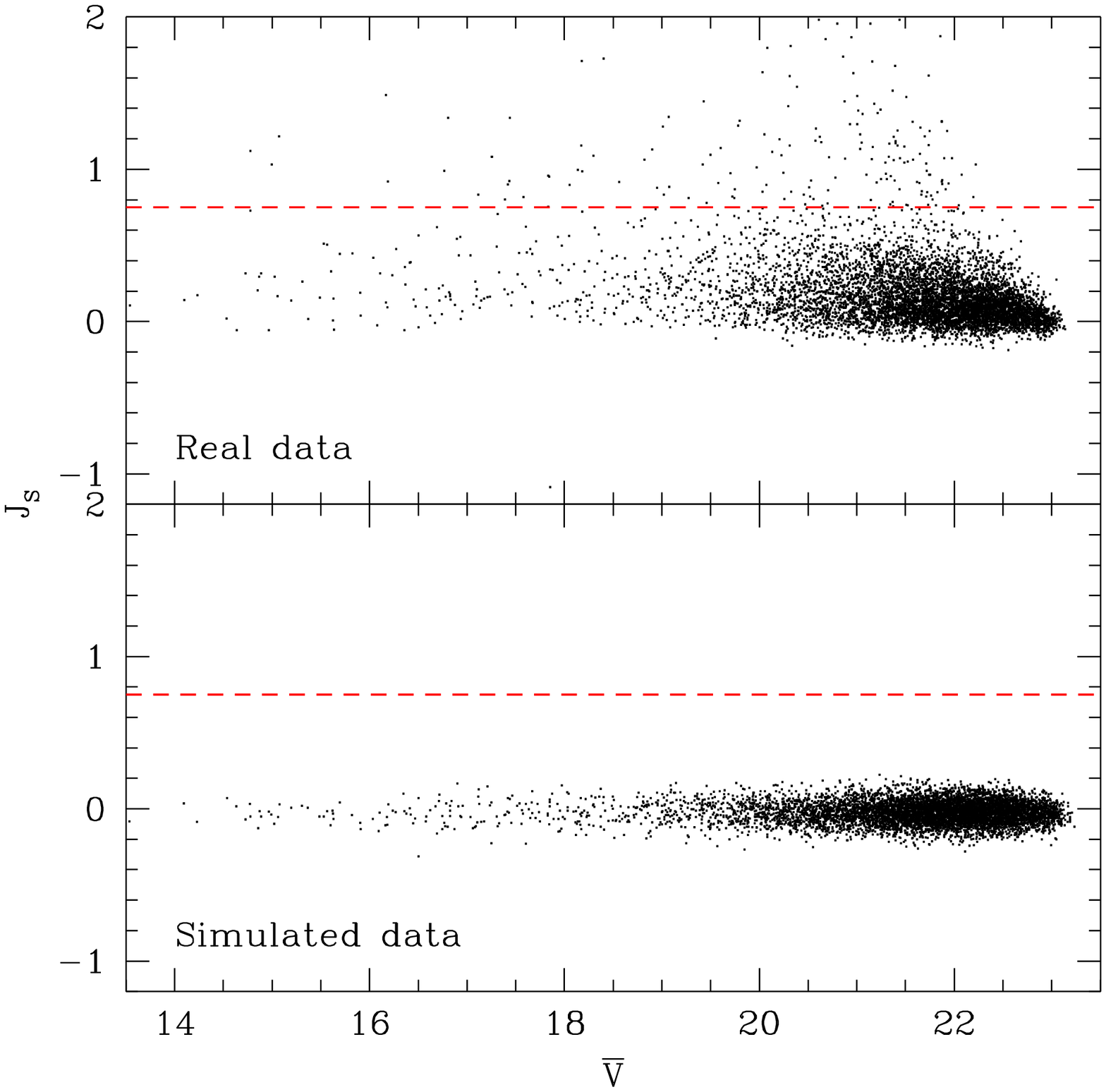}{8cm}{0}{50}{50}{-160}{-85}
\caption{Variability index $J_S$ vs. mean
$\bar{V}$ magnitude for stars with $N_{good}>80$, plotted for the real
data (upper panel) and the simulated Gaussian noise. In the case of
real data, there are stars with $J_S>2$ which are not shown. The
dashed line at $J_S=0.75$ defines the cutoff for variability we use.}
\label{fig:stetj}
\end{figure}

We selected the candidate variable stars by computing the value of
$J_S$ for the stars in our $V$ database.  In Fig.\ref{fig:stetj} we
plot the variability index $J_S$ vs. apparent visual magnitude
$\bar{V}$ for stars with $N_{good}>80$, for real data (upper panel)
and simulated Gaussian noise (lower panel). In the case of real data,
there are stars with $J_S>2$ which are not shown.  As expected (see
discussion in Stetson 1996), most of the stars have values of $J_S$
which are close to 0. Not surprisingly, the values of $J_s$ for the
real data are much more scattered, both due to the real variability,
as well as various un-modelled measurement errors.

We used a cutoff of $J_{S,min}=0.75$ to select 202 candidate variable
stars (about 3\% of the total number of 7208). There is one star with
abnormally {\em negative} value of $J_S$, located at
$(\bar{V},J_S)=(17.86,-1.09)$ in Fig.\ref{fig:stetj}, a contact binary
with period of $P=0.23\;day=5.5\;hour$ that is comparable to our
pairing interval of $1.5\;hour$. We decided to add this star to our
sample of candidate variables.

After a preliminary examination of the 203 candidate variables, we
decided to add two additional cuts. First, there are in our sample
many bright stars with variability of very small ($<0.03\;{\rm mag}$)
amplitude. The small variability might be real, since there are other
bright stars which show a random scatter of only $\sim 0.01\;{\rm
mag}$.  We decided, however, to remove variables for which the
standard deviation $\sigma$ of the magnitude measurements was smaller
than $\sigma<0.03\;{\rm mag}$. Second, we decided to remove from the
sample all the stars with the $x$ coordinate greater than
$x>2000$. Out of 56 stars with $x>2000$, 25 were classified as
variable ($J_S>0.75$), and the rest also had larger than normal values
of $J_S$. The anomalous properties are probably caused by especially
strong spatial variation of the PSF near this edge of the CCD. The
other edges of the CCD do not show such strong effect. We are left
with 163 candidate variable stars.

\section{Period determination, classification of variables}

\subsection{Additional data}

We based our candidate variables selection purely on the $V$ band data
collected at the MDM telescope. However, to better determine
the possible periods and to classify the variables, we added up to 6
$V$ band measurements taken at the FLWO telescope, which extended
the time span of observations for some stars to $56\;days$. 

We also have the $I$ band data for the field, up to 27 MDM epochs and
up to 17 FLWO epochs. As discussed earlier in this paper, the $I$
photometry is not as deep as the $V$ photometry, so some of the
candidate variable stars do not have an $I$ counterpart. We will
therefore not use the $I$ data for the period determination and broad
classification of the variables. We will however use the $I$ data for
the ``final'' classification of some variables.

\subsection{Period determination}

Next we searched for the periodicities for all 163 candidate
variables, using a variant of the Lafler-Kinman (1965) technique
proposed by Stetson (1996). Starting with the minimum period of
$0.25\;days$, successive trial periods are chosen so
\begin{equation}
P_{j+1}^{-1}=P_{j}^{-1}-\frac{0.01}{\Delta t},
\end{equation}
where $\Delta t=t_{N}-t_{1}$ is the time span of the series.  The
maximum period considered is $\Delta t$.  For each trial period, the
measurements are sorted by phase and we calculate
\begin{equation}
S(P)=\frac{\sum_{i=1}^{N}w(i,i+1)|m_i-m_{i+1}|}
{\sum_{i=1}^{N}w(i,i+1)},
\end{equation}
where
\begin{equation}
w(i,i+1)=\left[\frac{1}{\sigma_{i}^2+\sigma_{i+1}^2}\right].
\end{equation}
We did not use the additional phase difference weighting proposed by
Stetson (1996), because it tends to favor periods longer than the
``true'' period.  For all trial periods the values of $S(P)$ are
calculated, and 10 periods corresponding to the deepest local minima
of $S(P)$, separated from each other by at least $0.2/\Delta t$, are
selected. These 10 periods are then used in our classification scheme.

\subsection{Classification of variables}

The variables we are most interested in are Cepheids and eclipsing
binaries (EBs). We therefore searched our sample of variable stars for
these two classes of variables. As mentioned before, for the broad
classification of variables we restricted ourselves to the $V$ band
data.  We will, however, present and use the $I$ band data, when
available, when discussing some of the individual variable stars.

\subsubsection{Cepheid-like variables}

In the search for Cepheids we followed the approach by Stetson (1996)
of fitting template light curves to the data. We used the
parameterization of Cepheid light curves in the $V$ band as given by
Stetson (1996). Any template Cepheid light curve is determined by four
parameters: the period, the zero point of the phase, the amplitude and
the mean magnitude. From the template Cepheid we calculated the
expected magnitude of a Cepheid of the given parameters, and the
reduced $\chi^2/N_{DOF}$ for the fit of the model light curve to the
data. We minimize $\chi^2/N_{DOF}$ with a multidimensional
minimization routine.  We started the minimization with the ten best
trial periods from the Lafler-Kinman technique and we also used one
half of each value. After finding the best fit we classified the star
as a Cepheid if the reduced $\chi^2/N_{DOF}$ of the fit was factor of
2 smaller than the reduced $\chi^2/N_{DOF}$ of a straight line fit,
including a slope.  If a candidate satisfied these requirements we
restarted the minimization routine ten times with trial periods close
to the best fit period. Finally we required that the amplitude of the
best fit light curve was larger than $0.1\;{\rm mag}$.
 
\begin{figure}[t]
\plotfiddle{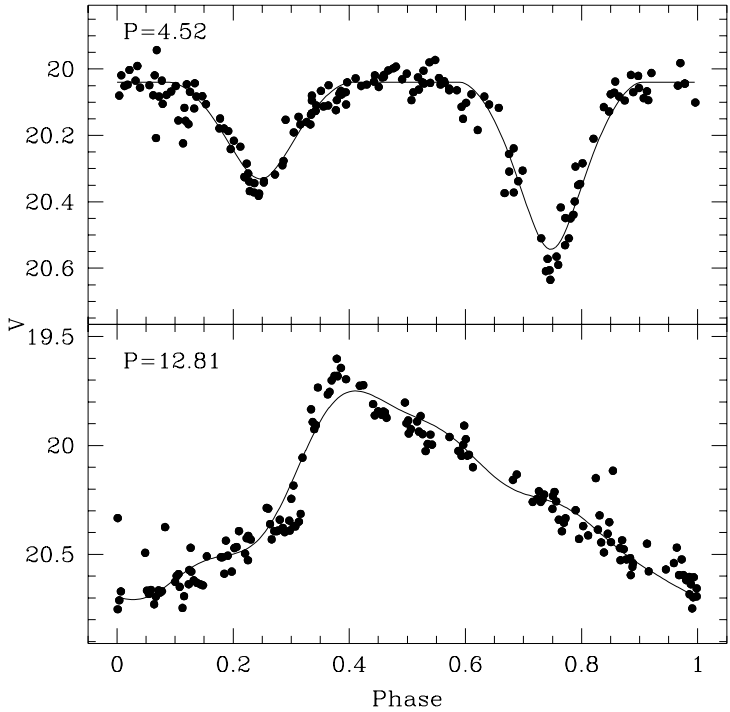}{8cm}{0}{50}{50}{-160}{-85}
\caption{Examples of fitting the model light curves to an eclipsing binary 
(upper panel) and to a Cepheid (lower panel). For the details see text.}
\label{fig:model}
\end{figure}

The template light curves we used were defined for period between 7
and $100\;days$, but we allowed for periods between 4 and $100\;days$.
The extension to smaller periods produced believable light curves.  As
can be seen in the lower panel of Fig.\ref{fig:model}, the fit of the
Cepheid template is not perfect: our data in this case is better than
what Stetson's templates were meant to fit (i.e.~sparsely sampled
Cepheid lightcurves obtained with the {\em HST}\/). However, for
purposes of discovery and period derivation these templates are
sufficient.

There was a total of 45 variables passing all of the above criteria.
Their parameters and light curves are presented in the
Sections~6.2,~6.3.

\subsubsection{Eclipsing binaries}

For eclipsing binaries we used very similar search strategy. We made
the simple assumption that the two stars in the binary system are
perfect spheres and have uniform surface brightnesses. This is not a
good assumption for detailed studies of the parameters of an EB, but
acceptable to calculate model light curves for trial fits. Within our
assumption the light curve of an EB is determined by nine parameters:
the period, the zero point of the phase, the eccentricity, the
longitude of periastron, the radii of the two stars relative to the
binary separation, the inclination angle, the fraction of light coming
from the bigger star and the uneclipsed magnitude. We derive the
orbital elements as a function of time by solving Kepler's equation.
A star was classified as an eclipsing variable if the reduced
$\chi^2/N_{DOF}$ of the EB light curve was smaller than the reduced
$\chi^2/N_{DOF}$ of a fit to a Cepheid and if it was smaller by a
factor of $4(R_1 + R_2) \sin^2 i$ than the reduced $\chi^2/N_{DOF}$ of
a fit to a line of constant magnitude. The $R_{1,2}$ are the radii of
the two stars in the binary relative to the binary separation and $i$
is the inclination angle. The scaling with the radii and the
inclination is necessary because some light curves show shallow and/or
narrow eclipses.  If a candidate star passed these two criteria we ran
the minimization routine ten more times with initial period guesses
close to the best fit period.  We required that the larger radius was
less than 0.9 of the binary separation and that the light of each
individual star was less than 0.9 of the total light. We further
rejected periods between 0.975 and 1.025 and between 1.95 and
$2.05\;days$. This last criterion was implemented to prevent us from
classifying as eclipsing binaries slowly varying stars, for which the
trial periods close to 1 and $2\;days$ produce spurious eclipsing-like
curves.

A total of 12 variables passing all of the above criteria and their
parameters and light curves are presented in the Section~6.1.  In the
upper panel of Fig.\ref{fig:model} we show an examples of fitting the
model light curves to an eclipsing binary.

\subsubsection{Miscellaneous variables}

After we selected 12 eclipsing binaries and 45 possible Cepheids, we
were left with 106 ``other'' variable stars. Visual inspection of
their phased and unphased light curves revealed both reasonably smooth
light curves as well as very chaotic or low amplitude
variability. Although we have already selected the variables we are
particularly interested in, it is of interest to others researchers to
present all highly probable variable stars in our data. We therefore
decided, for all variable stars other than Cepheids or eclipsing
binaries, to raise the threshold of the variability index to
$J_{S,min}=1.2$.  This leaves 37 variables which we preliminary
classify as ``miscellaneous''. One of these stars, V7453 D31B, was
clearly periodic, so we decided to classify it as ``other periodic
variable'' (see the Section~6.3). We then decided to go back to the
CCD frames and try to see by eye if the inferred miscellaneous
variability is indeed there, especially in cases when the light curve
is very noisy/chaotic.  This is obviously a rather subjective
procedure, and readers should employ caution when betting their life
savings on the reality of some of these candidates.  Note that we did
not apply this procedure to the eclipsing or Cepheid variables.

We decided to remove 9 dubious variables from the sample, which leaves
27 variables which we classify as miscellaneous. Their parameters and
light curves are presented in the Section~6.4.

\section{Catalog of variables}

In this section we present light curves and some discussion of the 85
variable stars discovered in our survey.  Complete $V$ and (when
available) $I$ photometry and $128\times128\;pixel$ ($\sim
40''\times40''$) $V$ finding charts for all variables are available
through the {\tt anonymous ftp} on {\tt cfa-ftp.harvard.edu}, in {\tt
pub/kstanek/DIRECT} directory. Please retrieve the {\tt README} file
for the instructions and the list of files. These data can also be
accessed through the {\tt WWW} at the {\tt
http://cfa-www.harvard.edu/\~\/kstanek/DIRECT/}.

The variable stars are named according to the following convention:
letter V for ``variable'', the number of the star in the $V$ database,
then the letter ``D'' for our project, DIRECT, followed by the name of
the field, in this case (M)31B, e.g. V888 D31B.
Tables~\ref{table:ecl}, \ref{table:ceph}, \ref{table:per} and
\ref{table:misc} list the variable stars sorted broadly by four
categories: eclipsing binaries, Cepheids, other periodic variables and
``miscellaneous'' variables, in our case meaning ``variables with no
clear periodicity''. Some of the variables which were found
independently by survey of Magnier et al.~(1997) are denoted in the
``Comments'' by ``Ma97 ID'', where the ``ID'' is the identification
number assigned by Magnier at al.~(1997).

Please note that this is a first step in a long-term project and we
are planning to collect additional data and information of various
kind for this and other fields we observed during 1996.  As a result,
the current catalog might undergo changes, due to addition, deletion
or re-classification of some variables. Please send an e-mail to
K. Z. Stanek ({\tt kstanek@cfa.harvard.edu}) if you want to be
informed of any such (major) changes.

\subsection{Eclipsing binaries}

In Table~\ref{table:ecl} we present the parameters of the 12 eclipsing
binaries in the M31B field.  The lightcurves of these variables are
shown in Figs.\ref{fig:ecl1}--\ref{fig:ecl2}, along with the simple
eclipsing binary models discussed in the Section~5.3.2.  The model
lightcurves were fitted to the $V$ data and then only a zero-point
offset was allowed for the $I$ data.  The variables are sorted in the
Table~\ref{table:ecl} by the increasing value of the period $P$. For
each eclipsing binary we present its name, 2000.0 coordinates (in
degrees), value of the variability index $J_S$, period $P$, magnitudes
$V_{max}$ and $I_{max}$ of the system outside of the eclipse, and the
radii of the binary components $R_1,\;R_2$ in the units of the orbital
separation.  We also give the inclination angle of the binary orbit to
the line of sight $i$ and the ellipticity of the orbit $e$. The reader
should bear in mind that the values of $V_{max},\;I_{max},\; R_1,\;R_2,\;i$
and $e$ are derived with a straightforward model of the eclipsing
system (Section 5.3.2), so they should be treated only as reasonable
estimates of the ``true'' value. As can be seen in Figs.\ref{fig:ecl1}
and \ref{fig:ecl2}, these simple binary models (shown with the thin
continuous lines) do a reasonable job in most of the cases.  More
detailed modeling will be performed of the follow-up observations
planned (see Section~7).

\begin{figure}[p]
\plotfiddle{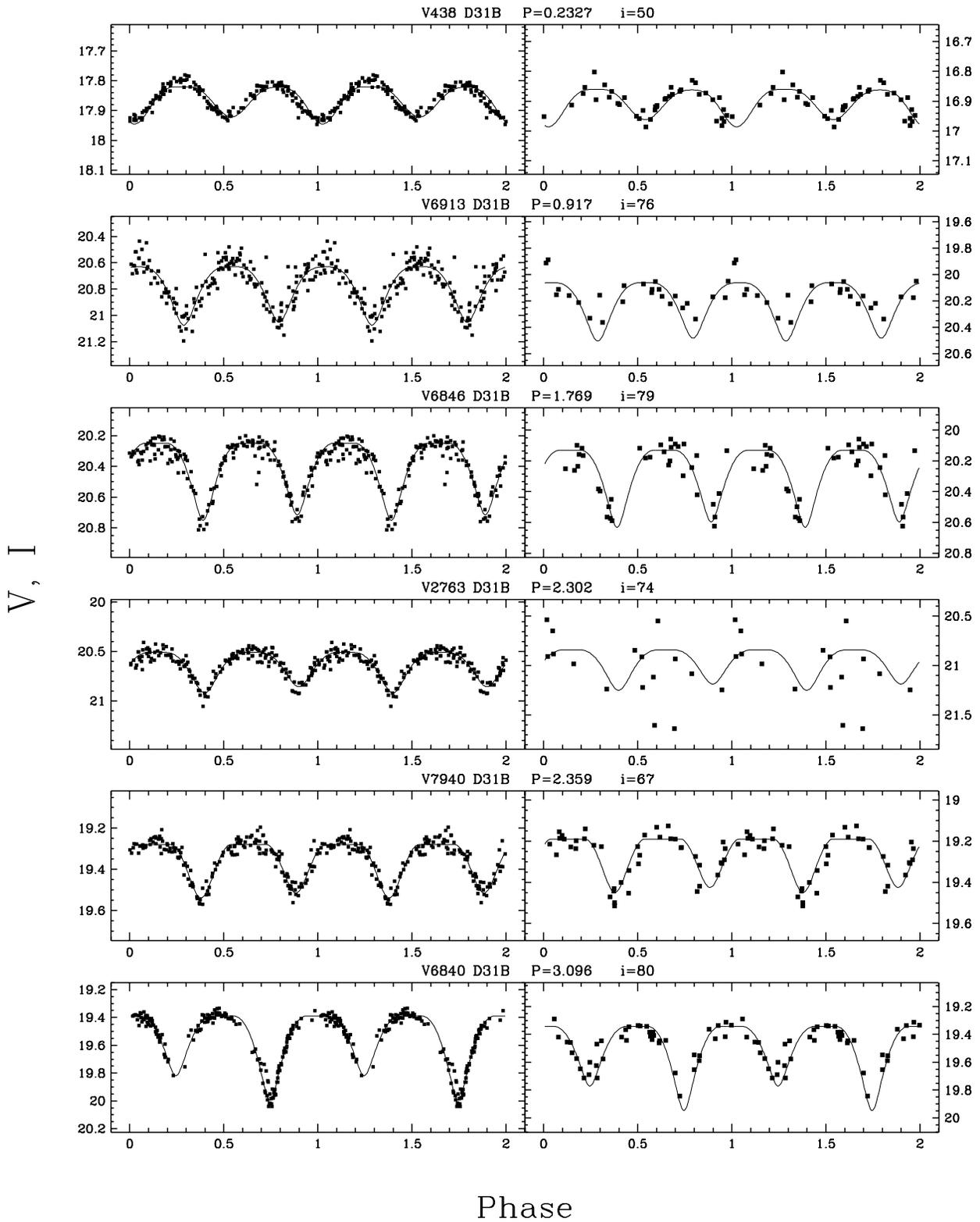}{19.5cm}{0}{83}{83}{-260}{-40}
\caption{$V,I$ lightcurves of eclipsing binaries found in the 
field M31B. The thin continuous line represents for each system
the best fit curve (fitted to the $V$ data).}
\label{fig:ecl1}
\end{figure}
\begin{figure}[p]
\plotfiddle{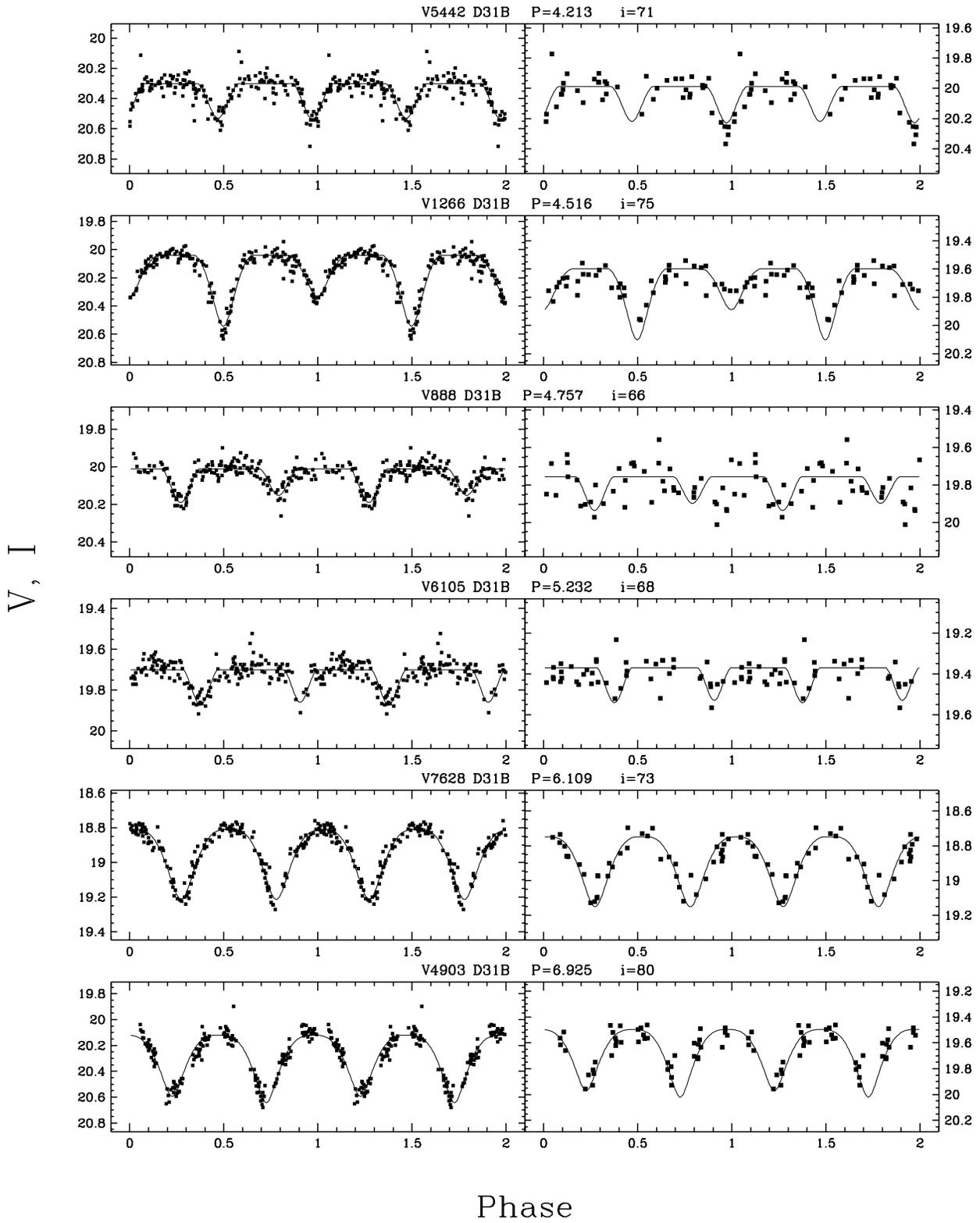}{19.5cm}{0}{83}{83}{-260}{-40}
\caption{Continued from  Fig.\ref{fig:ecl1}.}
\label{fig:ecl2}
\end{figure}

\subsection{Cepheids}

In Table~\ref{table:ceph} we present the parameters of 38 Cepheids in
the M31B field, sorted by the period $P$.  For each Cepheid we present
its name, 2000.0 coordinates, value of the variability index $J_S$,
period $P$, flux-weighted mean magnitudes $\langle V\rangle$ and (when
available) $\langle I\rangle$, and the amplitude of the variation
$A$. In Figs.\ref{fig:ceph1}--\ref{fig:ceph7} we show the phased $V,I$
lightcurves of our Cepheids. Also shown is the best fit template
lightcurve (Stetson 1996), which was fitted to the $V$ data and then
for the $I$ data only the zero-point offset was allowed.

\subsection{Other periodic variables}

For some of the variables preliminary classified as Cepheids
(Section~5.3.1), we decided upon closer examination to classify them
as ``other periodic variables''. This category includes also the
brightest variable star in the sample, V7453 D31B, which is a RR~Lyr
star. In Table~\ref{table:per} we present the parameters of 8 possible
periodic variables other than Cepheids and eclipsing binaries in the
M31B field, sorted by the increasing period $P$.  For each variable we
present its name, 2000.0 coordinates, value of the variability index
$J_S$, period $P$, error-weighted mean magnitudes $\bar{V}$ and (when
available) $\bar{I}$. To quantify the amplitude of the variability, we
also give the standard deviations of the measurements in the $V$ and
$I$ bands, $\sigma_{V}$ and $\sigma_{I}$.

\subsection{Miscellaneous  variables}

In Table~\ref{table:misc} we present the parameters of miscellaneous
variables in the M31B field, sorted by the decreasing value of the
mean magnitude $\bar{V}$. For each variable we present its name,
2000.0 coordinates, value of the variability index $J_S(>1.2)$, mean
magnitudes $\bar{V}$ and $\bar{I}$.  To quantify the amplitude of the
variability, we also give the standard deviations of the measurements
in $V$ and $I$ bands, $\sigma_{V}$ and $\sigma_{I}$.  In the
``Comments'' column we give a rather broad sub-classification of the
variability: LP -- possible long-period variable ($P>55\;days$); IRR
-- irregular variable.

\begin{figure}[p]
\plotfiddle{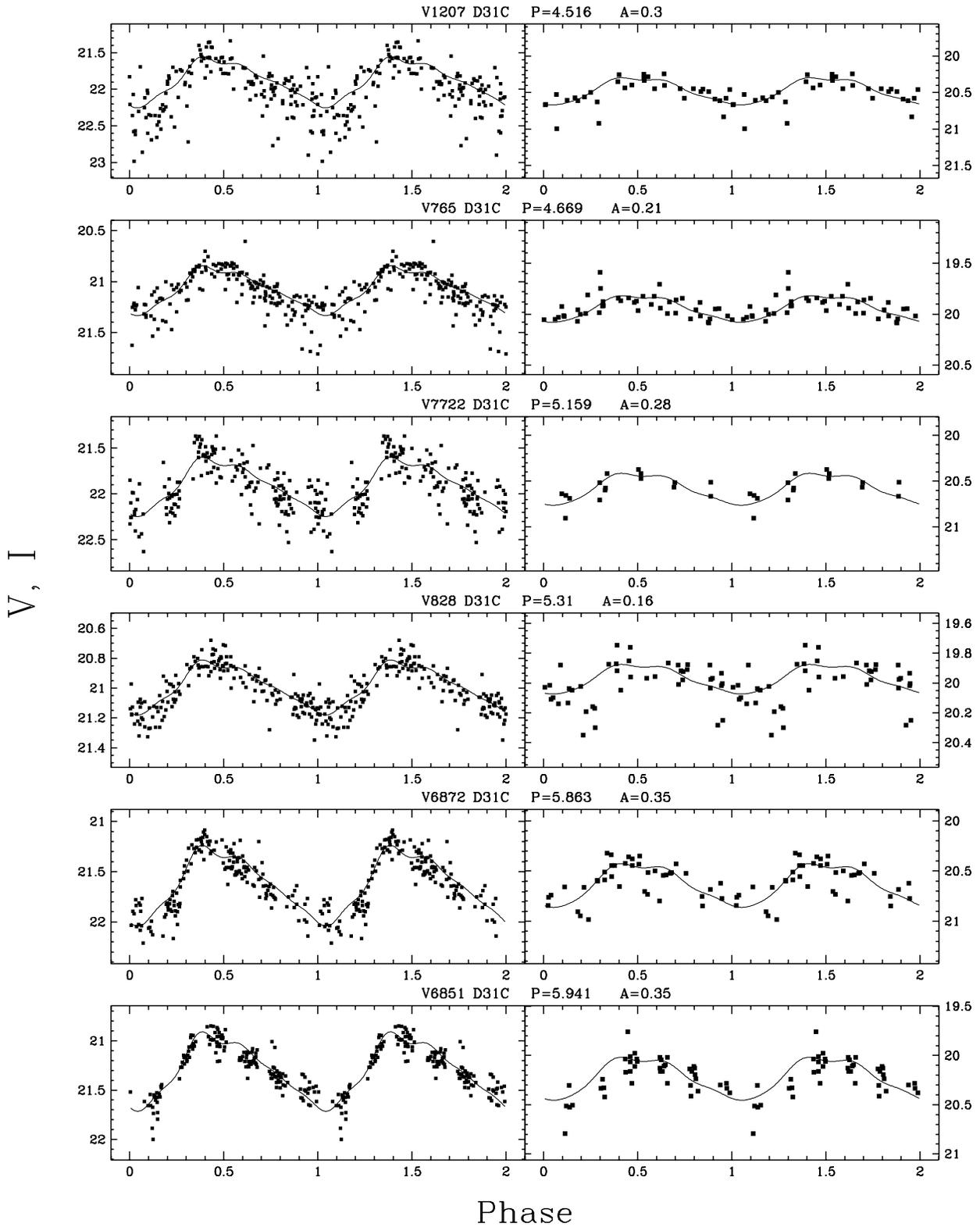}{19.5cm}{0}{83}{83}{-260}{-40}
\caption{$V,I$ lightcurves of Cepheid variables found in the 
field M31B. The thin continuous line represents for each star the best
fit Cepheid template (fitted to the $V$ data). }
\label{fig:ceph1}
\end{figure}
\begin{figure}[p]
\plotfiddle{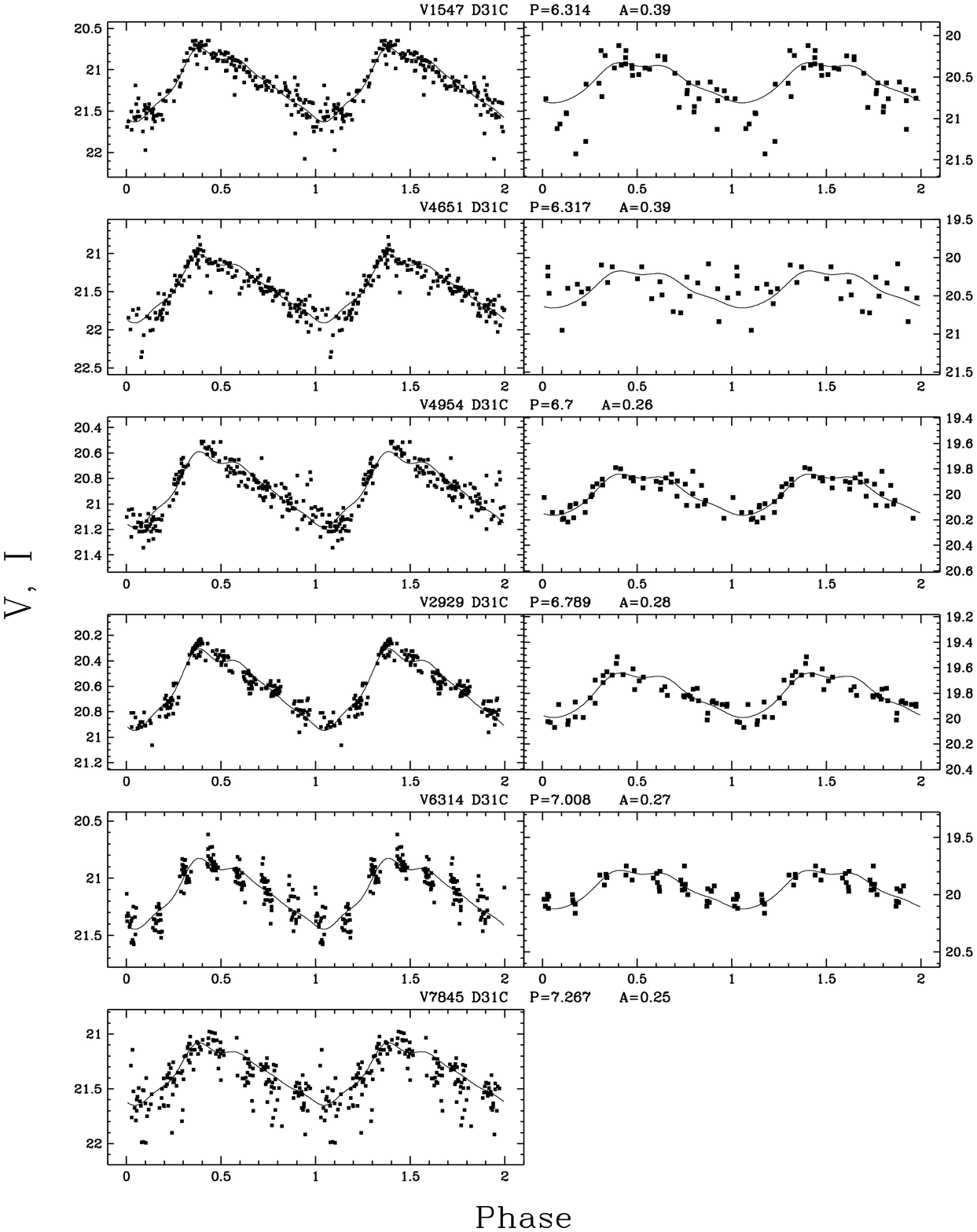}{19.5cm}{0}{83}{83}{-260}{-40}
\caption{Continued from  Fig.\ref{fig:ceph1}.}
\label{fig:ceph2}
\end{figure}
\begin{figure}[p]
\plotfiddle{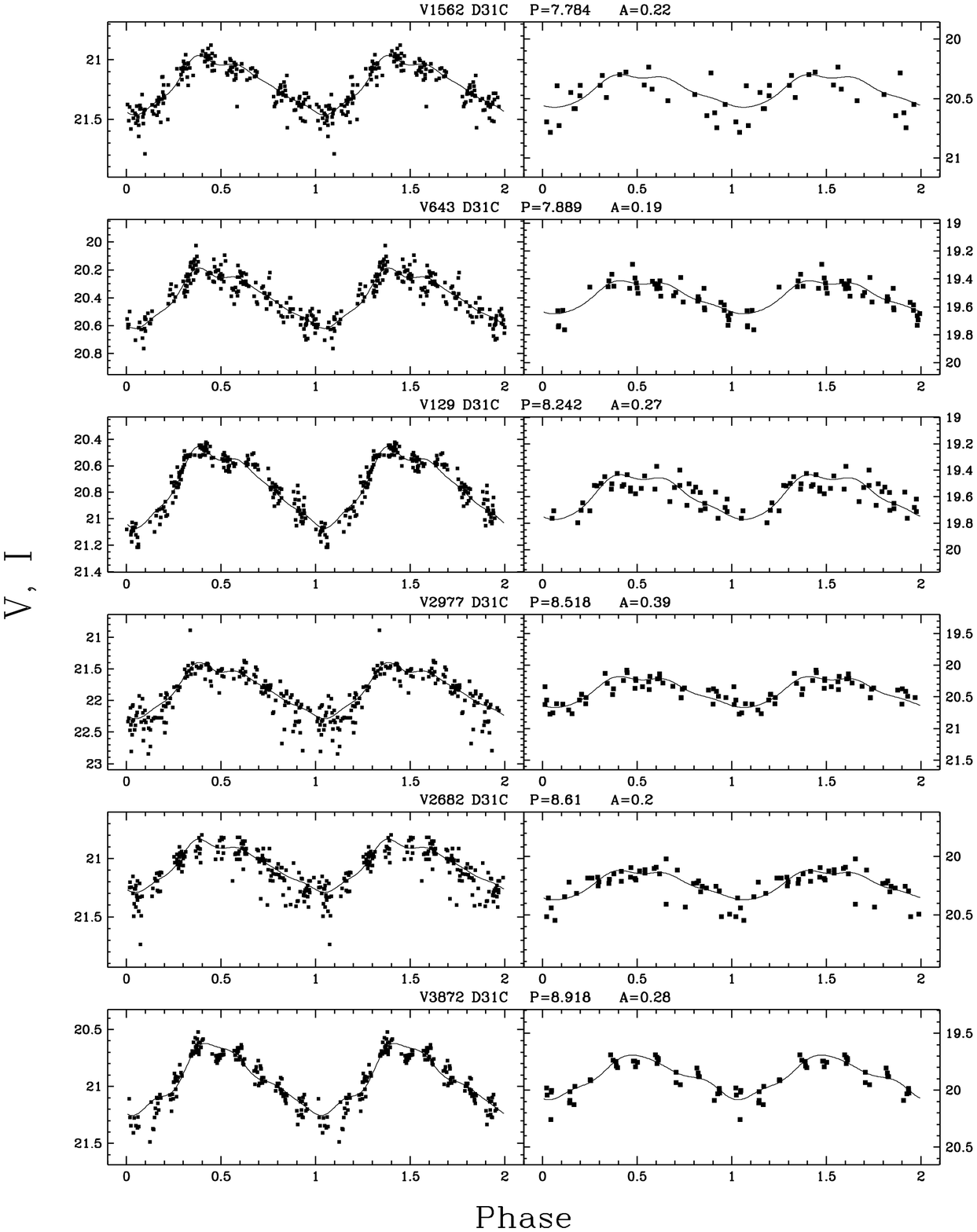}{19.5cm}{0}{83}{83}{-260}{-40}
\caption{Continued from  Fig.\ref{fig:ceph1}.}
\label{fig:ceph3}
\end{figure}
\begin{figure}[p]
\plotfiddle{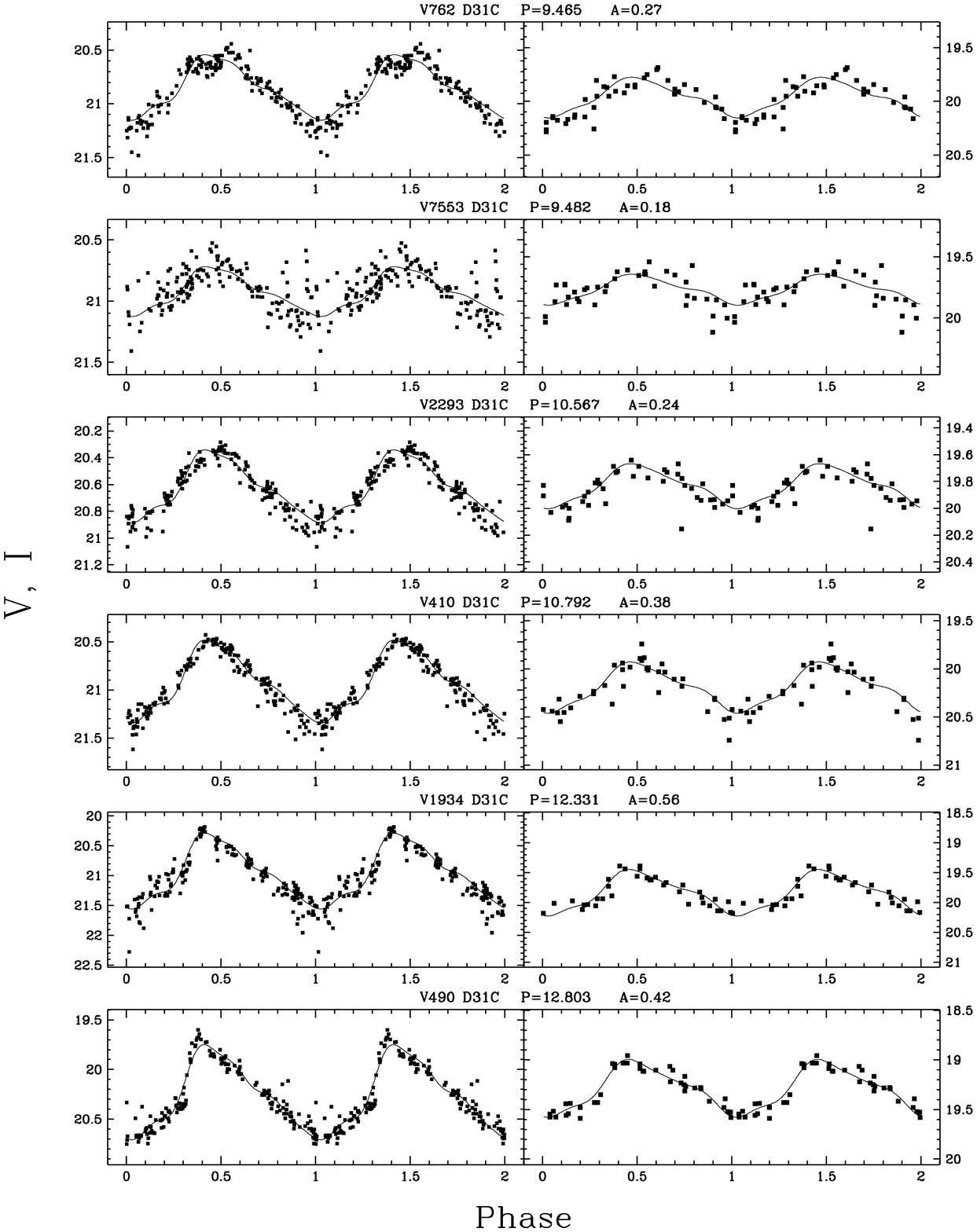}{19.5cm}{0}{83}{83}{-260}{-40}
\caption{Continued from  Fig.\ref{fig:ceph1}.}
\label{fig:ceph4}
\end{figure}
\begin{figure}[p]
\plotfiddle{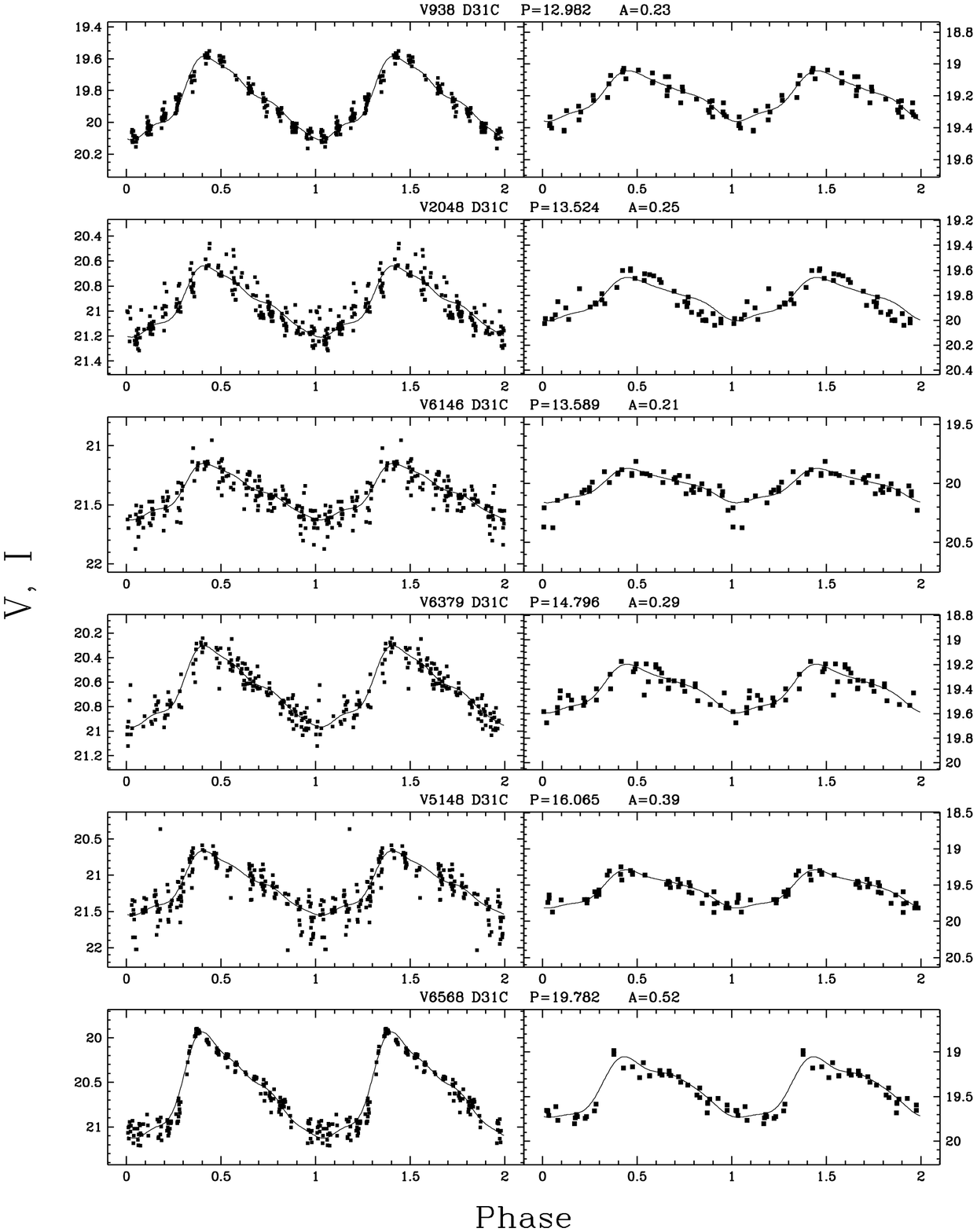}{19.5cm}{0}{83}{83}{-260}{-40}
\caption{Continued from  Fig.\ref{fig:ceph1}.}
\label{fig:ceph5}
\end{figure}
\begin{figure}[p]
\plotfiddle{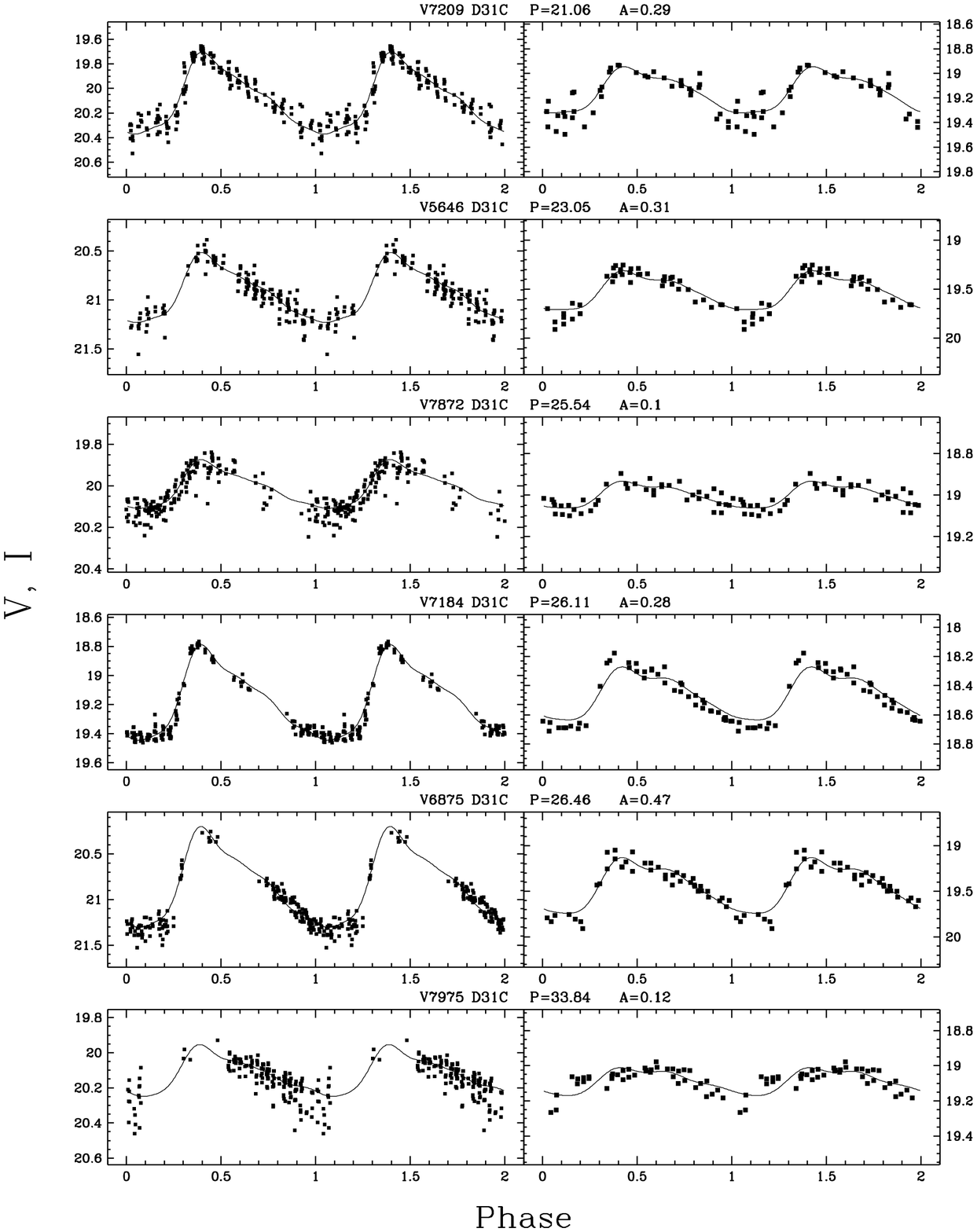}{19.5cm}{0}{83}{83}{-260}{-40}
\caption{Continued from  Fig.\ref{fig:ceph1}.}
\label{fig:ceph6}
\end{figure}
\begin{figure}[p]
\plotfiddle{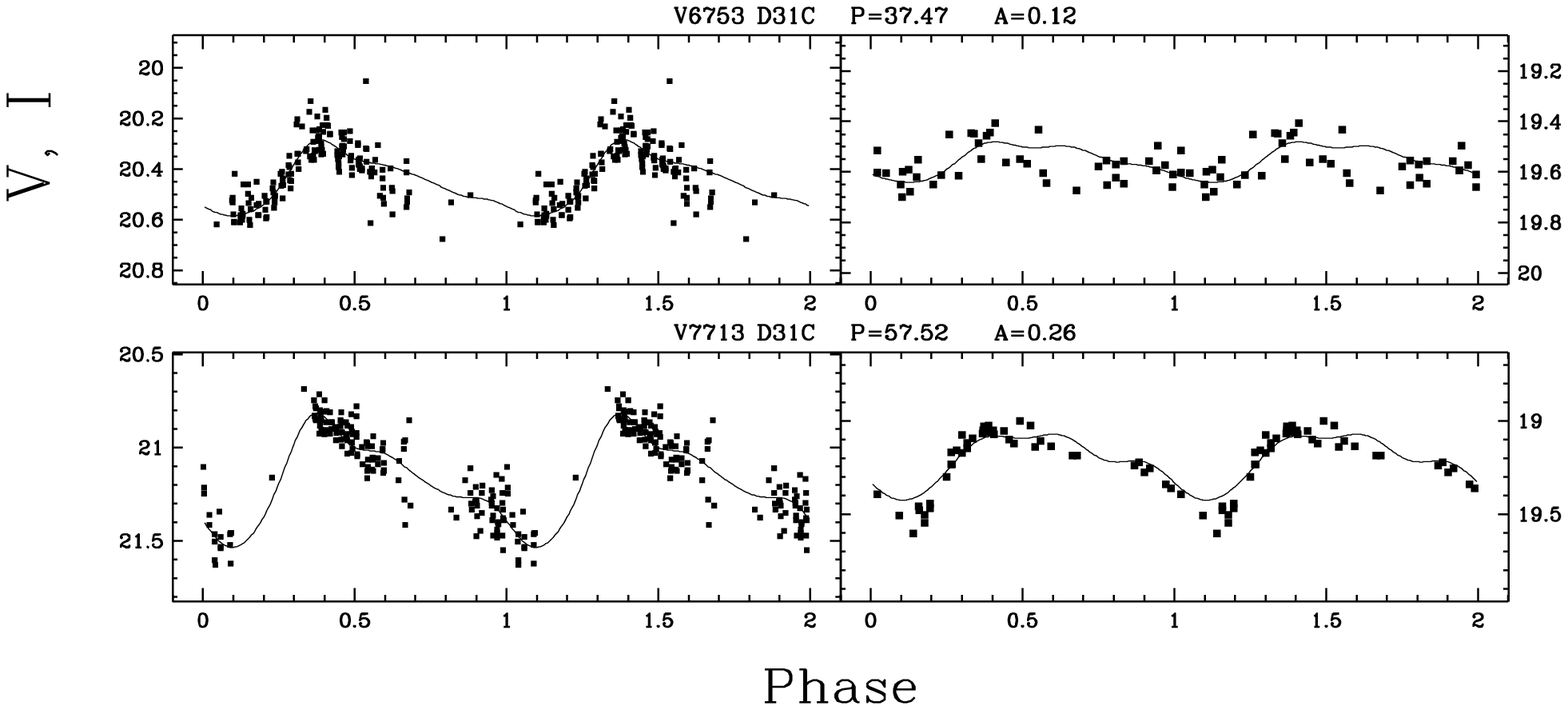}{6cm}{0}{83}{83}{-260}{-420}
\caption{Continued from  Fig.\ref{fig:ceph1}.}
\label{fig:ceph7}
\end{figure}

\begin{planotable}{rrrrrrrrrcrl}
\tablewidth{42pc}
\tablecaption{DIRECT Eclipsing Binaries in M31B}
\tablehead{ \colhead{Name} & \colhead{$\alpha_{2000.0}$} &
\colhead{$\delta_{2000.0}$} & \colhead{$J_S$} & \colhead{$P$}   &
\colhead{$V_{max}$} & \colhead{$I_{max}$} &
\colhead{$R_1$} & \colhead{$R_2$} & \colhead{$i$} & \colhead{$e$}
& \colhead{Comments} \\ \colhead{(D31B)} & \colhead{$[\deg]$} &
\colhead{$[\deg]$} & \colhead{} & \colhead{$[days]$} & \colhead{} & \colhead{} &
\colhead{} & \colhead{} & \colhead{[deg]}  & \colhead{} & \colhead{} } 
\startdata
 V438 & 11.0932 & 41.6475 &$-1.09$& ~0.2327 & 17.82 & 16.86 & 0.63 & 0.37 & 50 & 0.08 & W UMa  \nl
V6913 & 11.2717 & 41.6462 &  1.08 & 0.917~~ & 20.63 & 20.06 & 0.55 & 0.44 & 76 & 0.01 & \nl
V6846 & 11.2724 & 41.5612 &  1.61 & 1.769~~ & 20.25 & 20.13 & 0.51 & 0.42 & 79 & 0.00 & \nl
V2763 & 11.1568 & 41.4962 &  1.19 & 2.302~~ & 20.51 & 20.84 & 0.57 & 0.39 & 74 & 0.00 & \nl
V7940 & 11.3023 & 41.6240 &  2.08 & 2.359~~ & 19.28 & 19.19 & 0.51 & 0.34 & 67 & 0.01 & \nl
V6840 & 11.2703 & 41.6248 &  5.01 & 3.096~~ & 19.39 & 19.34 & 0.51 & 0.42 & 80 & 0.00 & \nl
V5442 & 11.2367 & 41.5197 &  0.85 & 4.213~~ & 20.30 & 19.99 & 0.39 & 0.29 & 71 & 0.02 & DEB \nl
V1266 & 11.1135 & 41.6023 &  1.80 & 4.516~~ & 20.04 & 19.60 & 0.48 & 0.36 & 75 & 0.00 & DEB? \nl
 V888 & 11.1033 & 41.6506 &  0.85 & 4.757~~ & 20.01 & 19.76 & 0.37 & 0.30 & 66 & 0.03 & DEB \nl
V6105 & 11.2520 & 41.5276 &  0.97 & 5.232~~ & 19.70 & 19.37 & 0.33 & 0.29 & 68 & 0.07 & DEB \nl
V7628 & 11.2930 & 41.6131 &  2.62 & 6.109~~ & 18.81 & 18.75 & 0.53 & 0.47 & 73 & 0.01 & \nl
V4903 & 11.2240 & 41.5196 &  2.40 & 6.925~~ & 20.12 & 19.50 & 0.56 & 0.44 & 80 & 0.00 & Ma97 96
\enddata 
\tablecomments{V438 D31B is most probably a foreground W UMa contact binary. 
V2763 D31B is very blue ($V-I\approx-0.3$), and the $I$ band data,
being very close to the detection limit, is very noisy. Variables
V5442, V1266, V888 and V6105, with periods from $P\approx4.2\;days$ to
$P\approx5.2\;days$, are probably detached eclipsing binaries (DEBs).}
\label{table:ecl}
\end{planotable}

\tablenum{2} 
\begin{planotable}{rrrrrrrrr}
\tablewidth{35pc}
\tablecaption{DIRECT Cepheids in M31B}
\tablehead{ \colhead{Name}  & \colhead{$\alpha_{2000.0}$} &
\colhead{$\delta_{2000.0}$} & \colhead{$J_S$} & \colhead{$P$}   &
\colhead{$\langle V\rangle$} & \colhead{$\langle I\rangle$} & \colhead{$A$} 
& \colhead{Comments} \\ \colhead{(D31B)} & \colhead{[deg]} &
\colhead{[deg]} & \colhead{} & \colhead{$[days]$} & \colhead{} &
\colhead{} & \colhead{} & \colhead{} } 
\startdata 
 V1207 & 11.1130 & 41.5680 & 0.96 &  4.516  & 21.89 & 20.47 & 0.30 & \nl
 V765  & 11.1022 & 41.6020 & 1.23 &  4.669  & 21.08 & 19.94 & 0.21 & \nl
 V7722 & 11.2972 & 41.5568 & 1.01 &  5.159  & 21.91 & 20.58 & 0.28 & \nl
 V828  & 11.1048 & 41.5640 & 1.31 &  5.310  & 20.99 & 19.97 & 0.16 & \nl
 V6872 & 11.2745 & 41.5124 & 1.55 &  5.863  & 21.63 & 20.63 & 0.35 & \nl
 V6851 & 11.2703 & 41.6342 & 1.56 &  5.941  & 21.30 & 20.22 & 0.35 & Ma97 106 \nl
 V1547 & 11.1194 & 41.6089 & 2.74 &  6.314  & 21.17 & 20.55 & 0.39 & \nl
 V4651 & 11.2146 & 41.5539 & 1.72 &  6.317  & 21.45 & 20.40 & 0.39 & \nl
 V4954 & 11.2250 & 41.5292 & 1.90 &  6.700  & 20.88 & 19.99 & 0.26 & Ma97 97 \nl
 V2929 & 11.1581 & 41.6269 & 2.84 &  6.789  & 20.62 & 19.81 & 0.28 & Ma97 87 \nl
 V6314 & 11.2544 & 41.6416 & 1.21 &  7.008  & 21.13 & 19.94 & 0.27 & \nl
 V7845 & 11.2983 & 41.6505 & 1.06 &  7.267  & 21.36 &\nodata& 0.25 & \nl
 V1562 & 11.1229 & 41.5087 & 1.23 &  7.784  & 21.21 & 20.43 & 0.22 & \nl
 V643  & 11.1021 & 41.5130 & 1.85 &  7.889  & 20.40 & 19.52 & 0.19 & \nl
 V129  & 11.0909 & 41.4971 & 2.27 &  8.242  & 20.76 & 19.59 & 0.27 & \nl
 V2977 & 11.1636 & 41.5022 & 1.59 &  8.518  & 21.84 & 20.41 & 0.39 & \nl
 V2682 & 11.1498 & 41.6212 & 1.37 &  8.610  & 21.06 & 20.24 & 0.20 & \nl
 V3872 & 11.1886 & 41.6339 & 2.14 &  8.918  & 20.95 & 19.88 & 0.28 & Ma97 93 \nl
 V762  & 11.1029 & 41.5792 & 2.27 &  9.465  & 20.86 & 19.96 & 0.27 & \nl
 V7553 & 11.2886 & 41.6657 & 1.11 &  9.482  & 20.93 & 19.77 & 0.18 & Ma97 108 \nl
 V2293 & 11.1385 & 41.6261 & 2.68 & 10.567  & 20.63 & 19.83 & 0.24 & Ma97 86 \nl
 V410  & 11.0918 & 41.6642 & 3.37 & 10.792  & 20.94 & 20.19 & 0.38 & \nl
 V1934 & 11.1291 & 41.6133 & 4.02 & 12.331  & 20.97 & 19.84 & 0.56 & \nl
 V490  & 11.0963 & 41.5883 & 5.01 & 12.803  & 20.27 & 19.29 & 0.42 & \nl
 V938  & 11.1078 & 41.5457 & 4.03 & 12.982  & 19.87 & 19.20 & 0.23 & Ma97 80 \nl
 V2048 & 11.1315 & 41.6321 & 1.76 & 13.524  & 20.95 & 19.83 & 0.25 & \nl
 V6146 & 11.2524 & 41.5531 & 0.99 & 13.589  & 21.41 & 20.02 & 0.21 &Ma97 102 \nl
 V6379 & 11.2585 & 41.5657 & 2.06 & 14.796  & 20.67 & 19.40 & 0.29 & Ma97 103 \nl
 V5148 & 11.2291 & 41.5531 & 2.19 & 16.065  & 21.16 & 19.57 & 0.39 & Ma97 98 \nl
 V6568 & 11.2635 & 41.6005 & 6.45 & 19.782  & 20.62 & 19.42 & 0.52 & Ma97 104 \nl
 V7209 & 11.2814 & 41.5877 & 3.46 & 21.06~\/ & 20.09 & 19.15 & 0.29 & \nl
 V5646 & 11.2414 & 41.5093 & 2.06 & 23.05~\/ & 20.93 & 19.52 & 0.31 & \nl
 V7872 & 11.2995 & 41.6419 & 1.69 & 25.54~\/ & 20.01 & 19.00 & 0.10 & \nl
 V7184 & 11.2797 & 41.6217 & 5.39 & 26.11~\/ & 19.17 & 18.46 & 0.28 & Ma97 107 \nl
 V6875 & 11.2739 & 41.5348 & 2.22 & 26.46~\/ & 20.84 & 19.45 & 0.47 & \nl
 V7975 & 11.3061 & 41.5349 & 1.12 & 33.84~\/ & 20.12 & 19.09 & 0.12 & \nl
 V6753 & 11.2710 & 41.5228 & 1.76 & 37.47~\/ & 20.45 & 19.55 & 0.12 &\nl
 V7713 & 11.2957 & 41.6041 & 1.82 & 57.52~\/ & 21.18 & 19.21 & 0.26 & 
\enddata
\label{table:ceph}
\end{planotable}

\tablenum{3} 
\begin{planotable}{lllrrlllll}
\tablewidth{38pc}
\tablecaption{DIRECT Other Periodic Variables in M31B}
\tablehead{ \colhead{Name} & \colhead{$\alpha_{2000.0}$} &
\colhead{$\delta_{2000.0}$} & \colhead{$J_S$} & \colhead{$P$}   &
\colhead{$\bar{V}$} & \colhead{$\bar{I}$} &
\colhead{$\sigma_V$} & \colhead{$\sigma_I$} & \colhead{Comments} \\
\colhead{(D31B)} &  \colhead{[deg]} &  \colhead{[deg]} &
\colhead{} & \colhead{$[days]$} & \colhead{} &
\colhead{} & \colhead{} & \colhead{} & \colhead{} }
\startdata
V7453  & 11.2909 & 41.5086 & 16.99 &  ~0.579   & 16.72 & 16.28 & 0.26 & 0.16  & RR Lyr \nl
V6518  & 11.2610 & 41.6219 &  0.97 &  ~9.686   & 21.10 & 20.61 & 0.16 & 0.20  & \nl
V3825  & 11.1899 & 41.5375 &  1.16 &  22.02~   & 21.44 & 20.44 & 0.28 & 0.25  & W Vir? \nl
V7341  & 11.2835 & 41.6406 &  0.77 &  29.00~   & 21.48 &\nodata& 0.25 &\nodata& \nl
V4773  & 11.2197 & 41.5005 &  2.06 &  30.26~   & 21.13 & 20.28 & 0.36 & 0.19  & RV Tau \nl
V6164  & 11.2506 & 41.6245 &  2.49 &  32.72~   & 21.44 & 20.94 & 0.53 & 0.44  & RV Tau? \nl
V1290  & 11.1135 & 41.6162 &  0.76 &  44.7~~\/ & 21.38 & 20.42 & 0.18 & 0.11  & RV Tau? \nl
V3469  & 11.1783 & 41.5376 &  1.25 &  48.0~~\/ & 21.68 & 20.74 & 0.43 & 0.38  & 
\enddata								
\label{table:per}							
\end{planotable}							

\tablenum{4} 
\begin{planotable}{lllllllll}
\tablewidth{35pc}
\tablecaption{DIRECT Miscellaneous Variables in M31B}
\tablehead{ \colhead{Name} & \colhead{$\alpha_{2000.0}$} &
\colhead{$\delta_{2000.0}$} & \colhead{$J_S$} &
\colhead{$\bar{V}$} & \colhead{$\bar{I}$} &
\colhead{$\sigma_V$} & \colhead{$\sigma_I$} & \colhead{Comments} \\
\colhead{(D31B)} & \colhead{[deg]} & \colhead{[deg]} &
\colhead{} & \colhead{} & \colhead{} &
\colhead{} & \colhead{} & \colhead{} }
\startdata
V5688 & 11.2417 & 41.5304 & 2.34 & 17.55 & 16.81 & 0.05 & 0.03 & Foreground  \nl
V4697 & 11.2172 & 41.5069 & 2.31 & 18.13 & 17.55 & 0.05 & 0.11 & Foreground  \nl
V6496 & 11.2638 & 41.5072 & 1.71 & 18.20 & 15.96 & 0.04 & 0.03 & LP  \nl
V7984 & 11.3061 & 41.5425 & 4.66 & 18.20 & 17.09 & 0.10 & 0.08 & IRR \nl
V7606 & 11.2911 & 41.6454 & 1.28 & 19.03 & 16.81 & 0.06 & 0.03 & LP  \nl
V5830 & 11.2449 & 41.5139 & 2.02 & 19.07 & 16.75 & 0.08 & 0.04 & LP  \nl
V8123 & 11.3081 & 41.6494 & 2.42 & 19.12 & 17.27 & 0.10 & 0.07 & LP  \nl
V1019 & 11.1073 & 41.6165 & 2.08 & 19.17 & 17.41 & 0.09 & 0.06 & LP  \nl
V8197 & 11.3121 & 41.6263 & 1.45 & 19.45 & 16.35 & 0.08 & 0.04 & LP  \nl
V4062 & 11.1987 & 41.5138 & 3.60 & 19.75 & 16.72 & 0.13 & 0.05 & LP  \nl
V7326 &	11.2839	& 41.6152 & 1.29 & 19.77 & 17.13 & 0.07	& 0.03 & LP  \nl
V6936 & 11.2726 & 41.6354 & 1.32 & 19.80 & 19.38 & 0.08 & 0.09 & IRR \nl
V7797 &	11.2979	& 41.6375 & 1.23 & 20.04 & 20.57 & 0.09	& 0.13 & blue \nl
V5724 & 11.2396 & 41.6256 & 1.41 & 20.31 & 19.41 & 0.20 & 0.09 & LP  \nl
V3333 & 11.1717 & 41.6167 & 4.06 & 20.82 & 20.23 & 0.64 & 0.41 & LP  \nl
V5504 & 11.2382 & 41.5095 & 1.30 & 20.95 & 18.52 & 0.19 & 0.17 & LP  \nl
V3237 & 11.1698 & 41.5627 & 2.03 & 20.99 & 20.23 & 0.24 & 0.35 & Ma97 89\nl
V6222 & 11.2511 & 41.6549 & 1.38 & 21.04 & 20.02 & 0.27 & 0.16 & LP  \nl
V4309 & 11.2062 & 41.5118 & 1.39 & 21.29 & 19.80 & 0.63 & 0.17 & LP  \nl
V4719 & 11.2169 & 41.5402 & 1.52 & 21.39 & 20.30 & 0.45 & 0.13 & LP  \nl
V4669 &	11.2920	& 41.5467 & 1.22 & 21.40 & 19.48 & 0.27	& 0.16 & LP  \nl
V5897 & 11.2424 & 41.6594 & 1.98 & 21.56 & 19.22 & 0.54 & 0.16 & LP  \nl
V7745 & 11.2991 & 41.5104 & 2.01 & 21.72 & 19.60 & 0.41 & 0.11 & LP  \nl
V2356 &	11.1412	& 41.5966 & 1.22 & 21.73 & 19.53 & 0.54	& 0.17 & LP  \nl
V7746 & 11.2982 & 41.5416 & 1.31 & 21.83 & 19.92 & 0.38 & 0.17 & LP  \nl
V5075 & 11.2239 & 41.6504 & 1.31 & 21.85 & 19.31 & 0.48 & 0.09 & LP  \nl
V4690 & 11.2142 & 41.5988 & 1.88 & 21.90 & 19.78 & 0.48 & 0.14 & LP 
\enddata
\label{table:misc}
\end{planotable}

\begin{figure}[p]
\plotfiddle{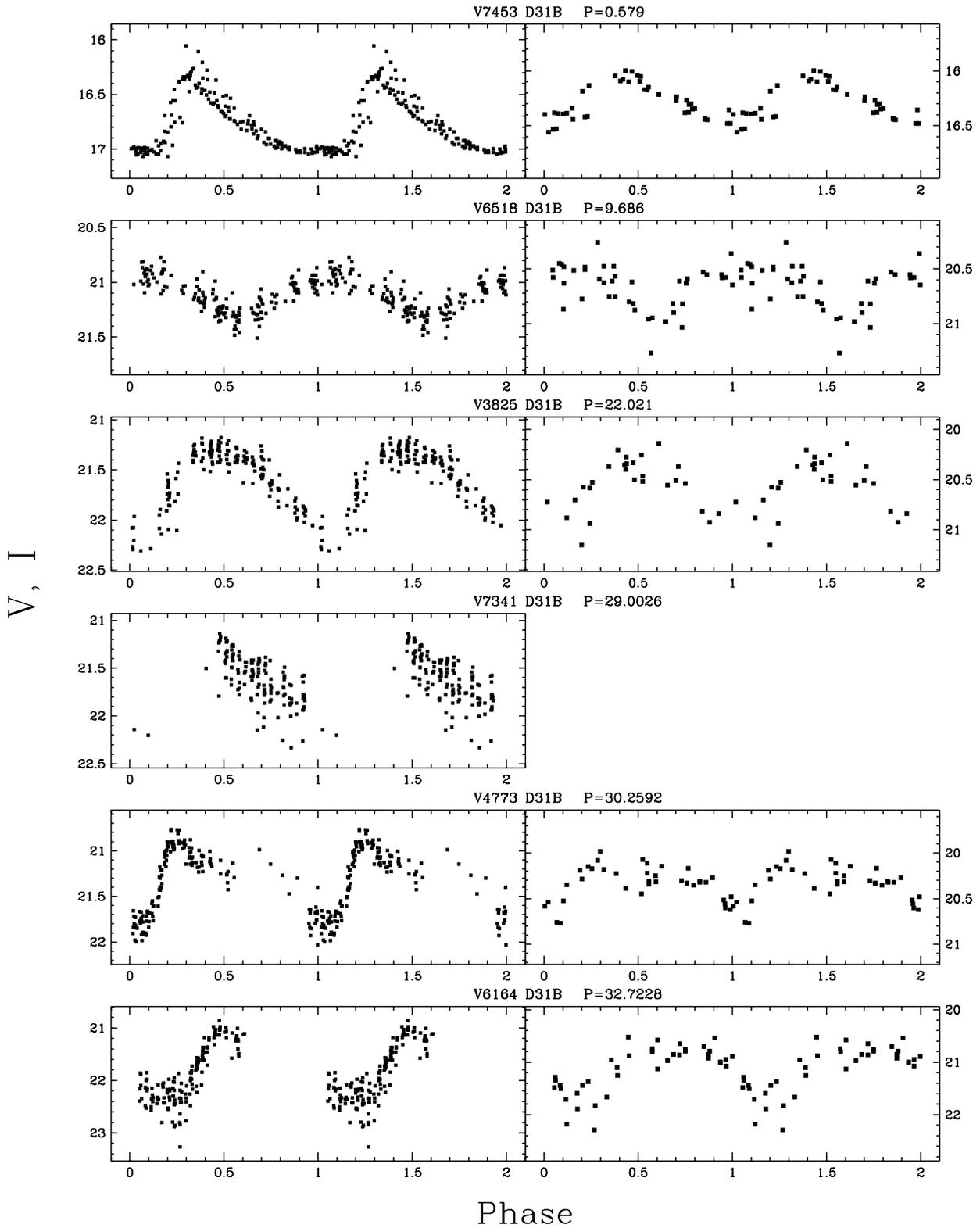}{19.5cm}{0}{83}{83}{-260}{-40}
\caption{$V,I$ lightcurves of other periodic  variables found in the 
field M31B.}
\label{fig:per1}
\end{figure}
\begin{figure}[p]
\plotfiddle{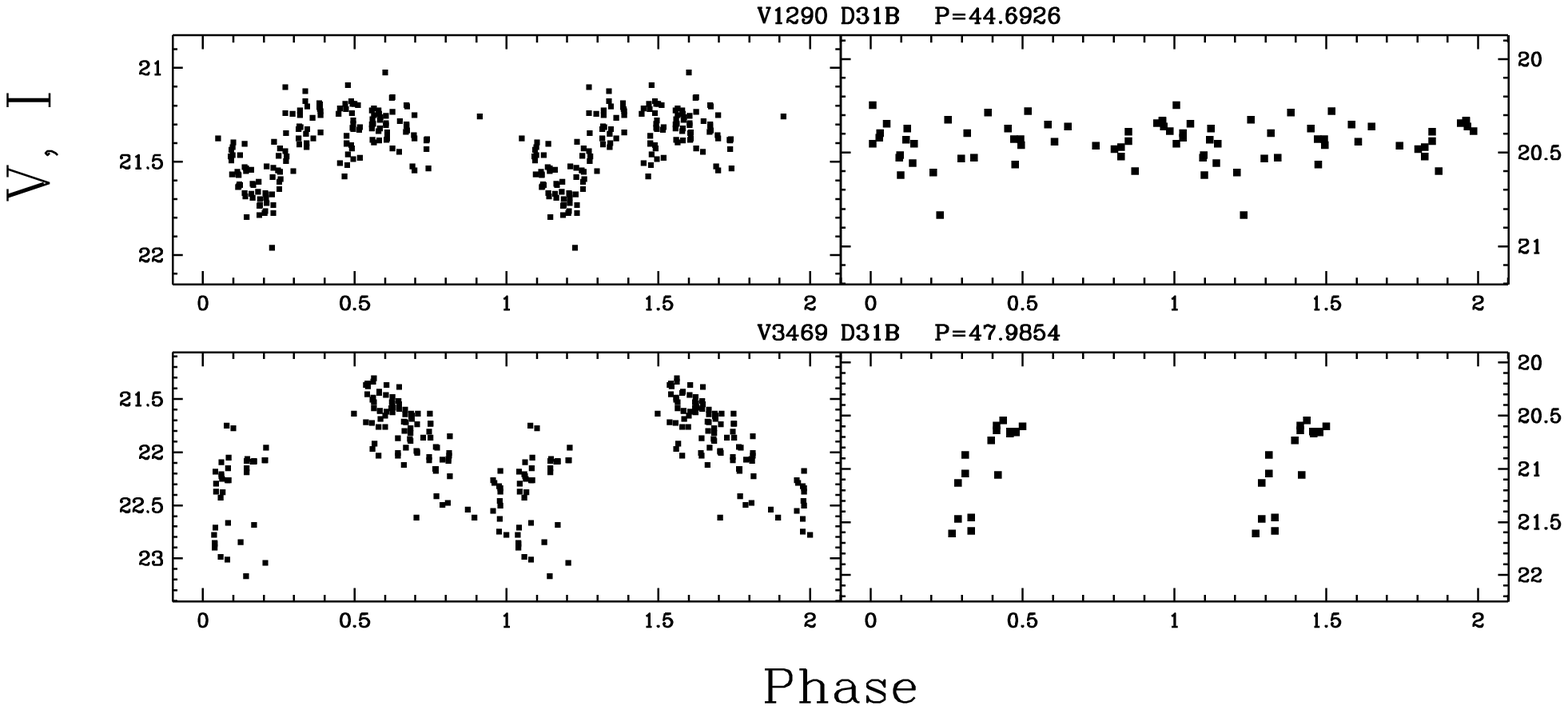}{6cm}{0}{83}{83}{-260}{-420}
\caption{Continued from  Fig.\ref{fig:per1}.}
\label{fig:per2}
\end{figure}

\begin{figure}[p]
\plotfiddle{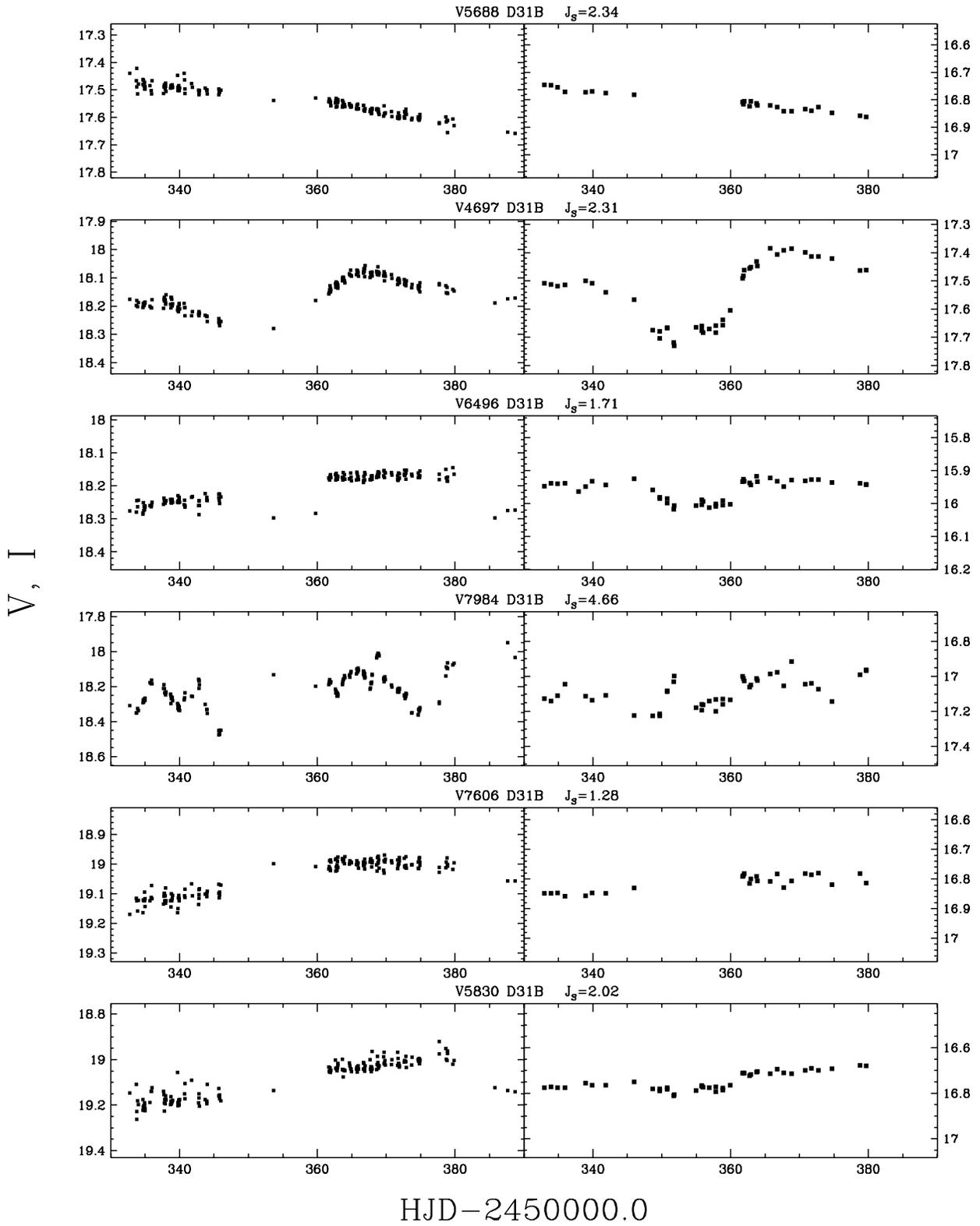}{19.5cm}{0}{83}{83}{-260}{-40}
\caption{$V,I$ lightcurves of miscellaneous variables found in the 
field M31B.}
\label{fig:misc1}
\end{figure}
\begin{figure}[p]
\plotfiddle{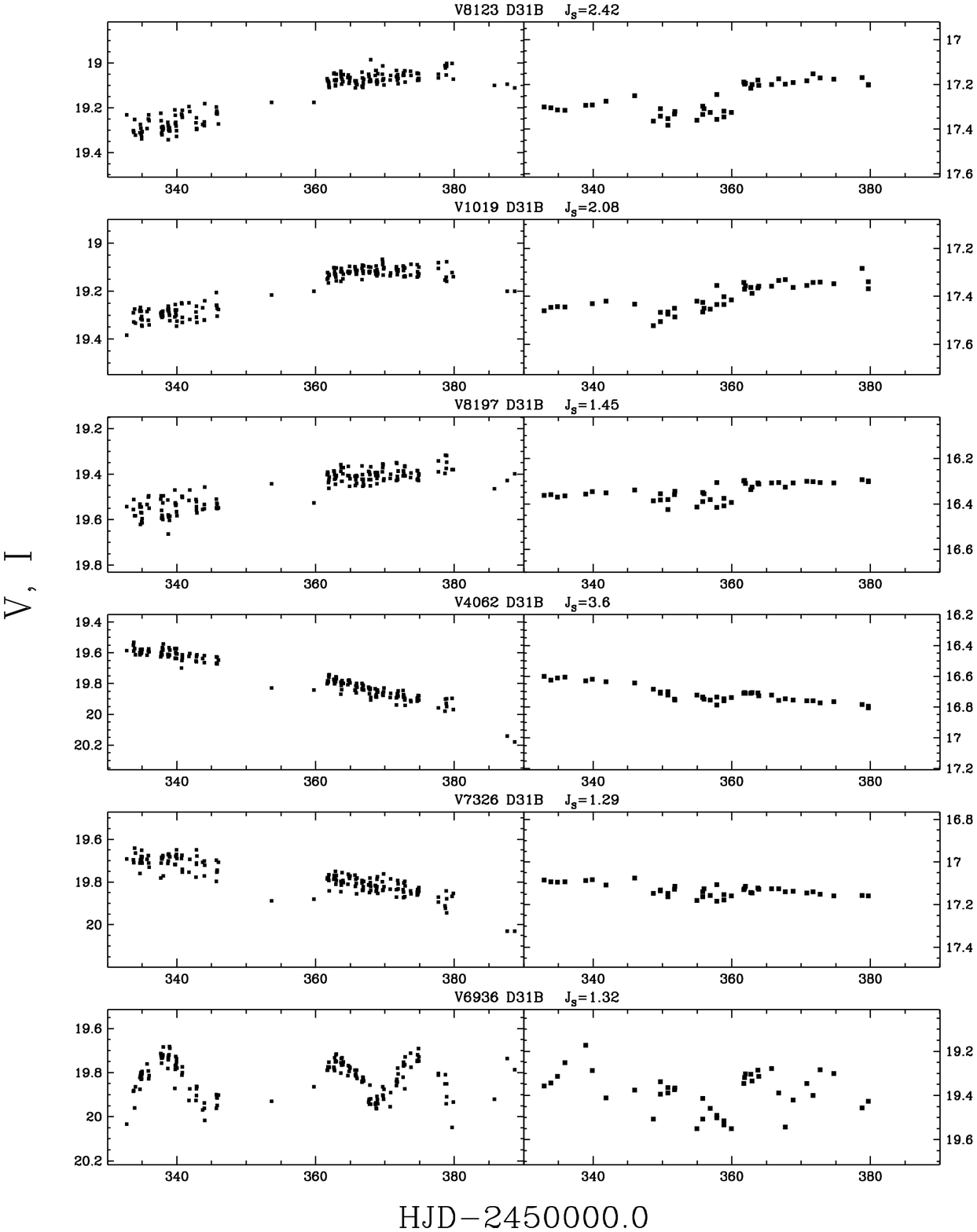}{19.5cm}{0}{83}{83}{-260}{-40}
\caption{Continued from  Fig.\ref{fig:misc1}.}
\label{fig:misc2}
\end{figure}
\begin{figure}[p]
\plotfiddle{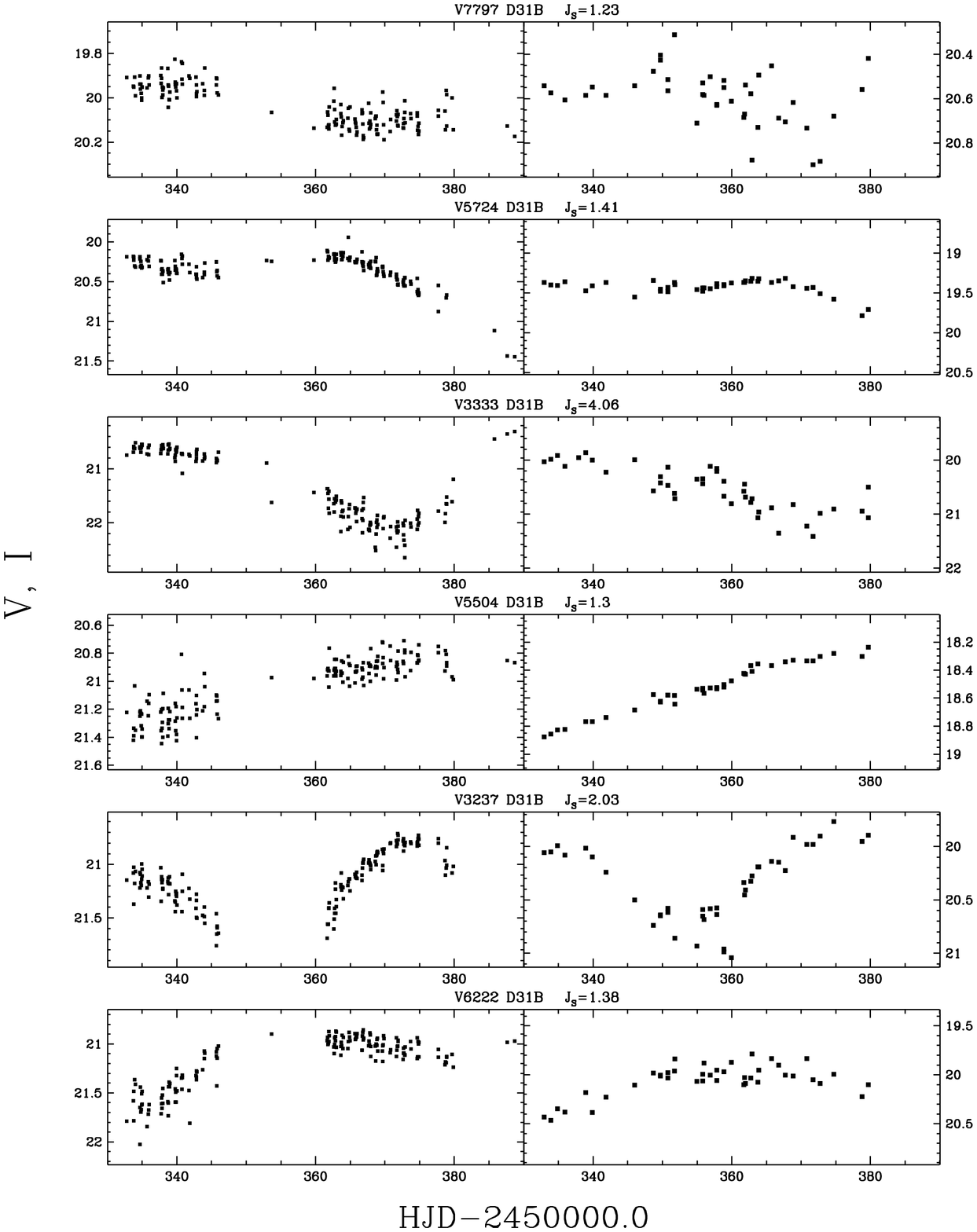}{19.5cm}{0}{83}{83}{-260}{-40}
\caption{Continued from  Fig.\ref{fig:misc1}.}
\label{fig:misc3}
\end{figure}
\begin{figure}[p]
\plotfiddle{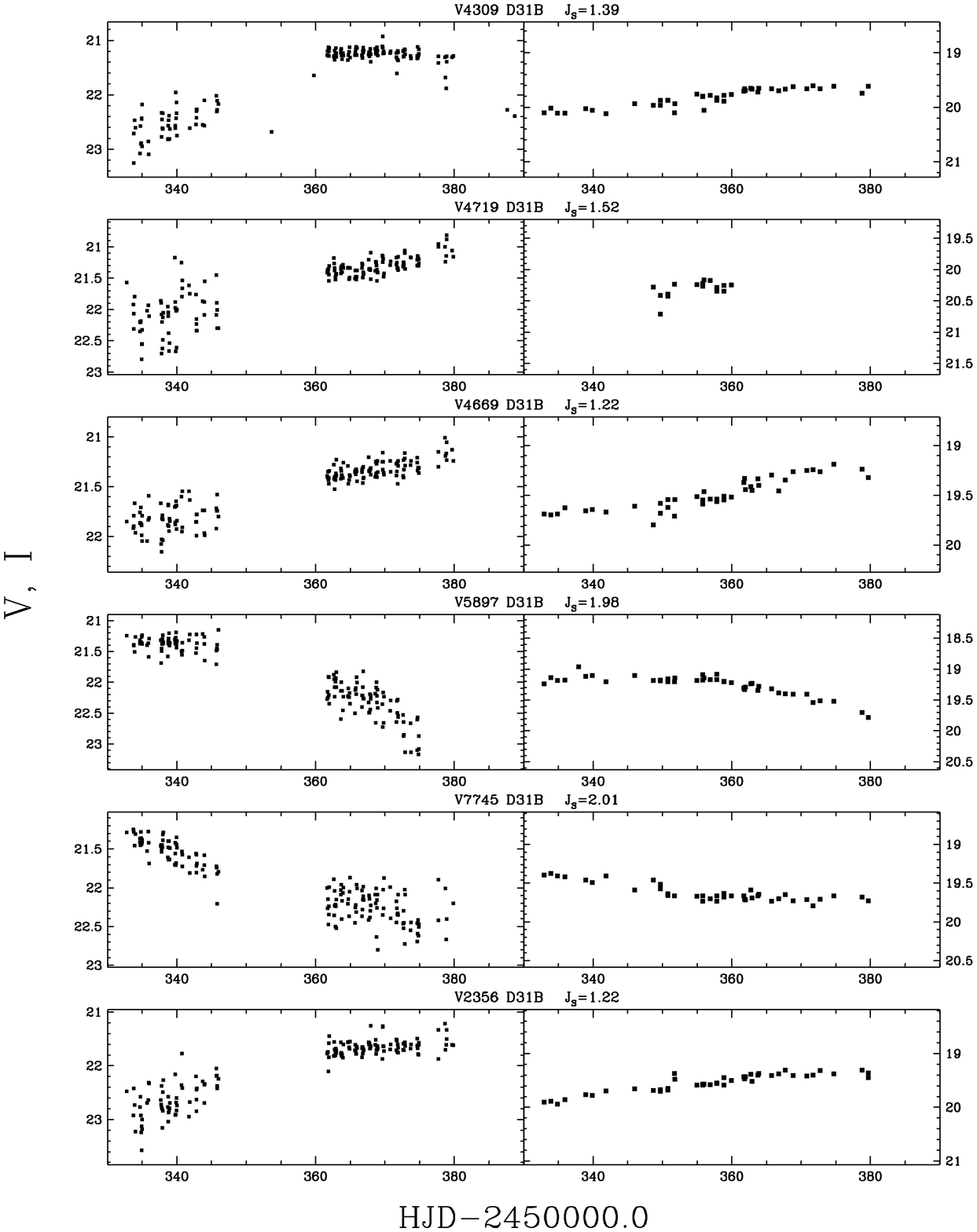}{19.5cm}{0}{83}{83}{-260}{-40}
\caption{Continued from  Fig.\ref{fig:misc1}.}
\label{fig:misc4}
\end{figure}
\begin{figure}[t]
\plotfiddle{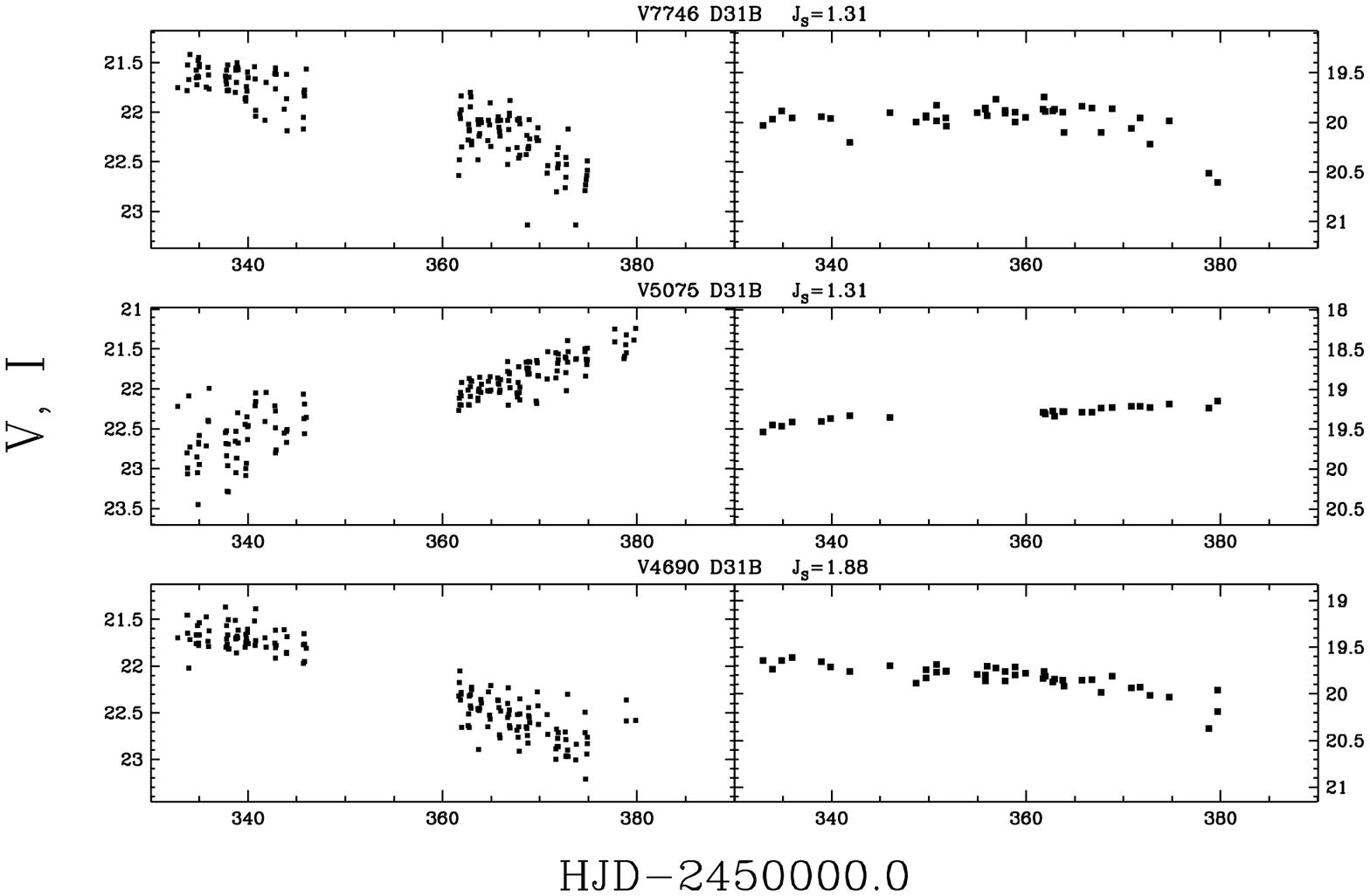}{9.3cm}{0}{83}{83}{-260}{-330}
\caption{Continued from  Fig.\ref{fig:misc1}.}
\label{fig:misc5}
\end{figure}

\subsection{Comparison with other catalogs}

The M31B field has not been observed frequently before and the only
overlapping variable star catalog is given by Magnier et al.~(1997,
hereafter Ma97). Of 16 variable stars in Ma97 which are in our M31B
field, we cross-identified 15. Of these 15 stars, one (Ma97 101) we
did not classify as a variable ($J_S=0.49$), one (V4903 D31B = Ma97
96) was classified as an eclipsing binary and one (V3237 D31B = Ma97
89) we classified as miscellaneous variable. The remaining 12 stars we
classified as Cepheids (see Table~\ref{table:ceph} for
cross-identifications).  Our M31B field also includes a confirmed
Luminous Blue Variable (see Humpreys \& Davidson 1994 for a review),
M31 Var A-1 (Humpreys 1997, private communication). We
cross-identified M31~Var~A-1 in our data and found it to be
non-variable, with average magnitudes $\bar{V}=16.80,\;
\bar{I}=16.10$.

\section{Discussion, follow-up observations and research}

\begin{figure}[t]
\plotfiddle{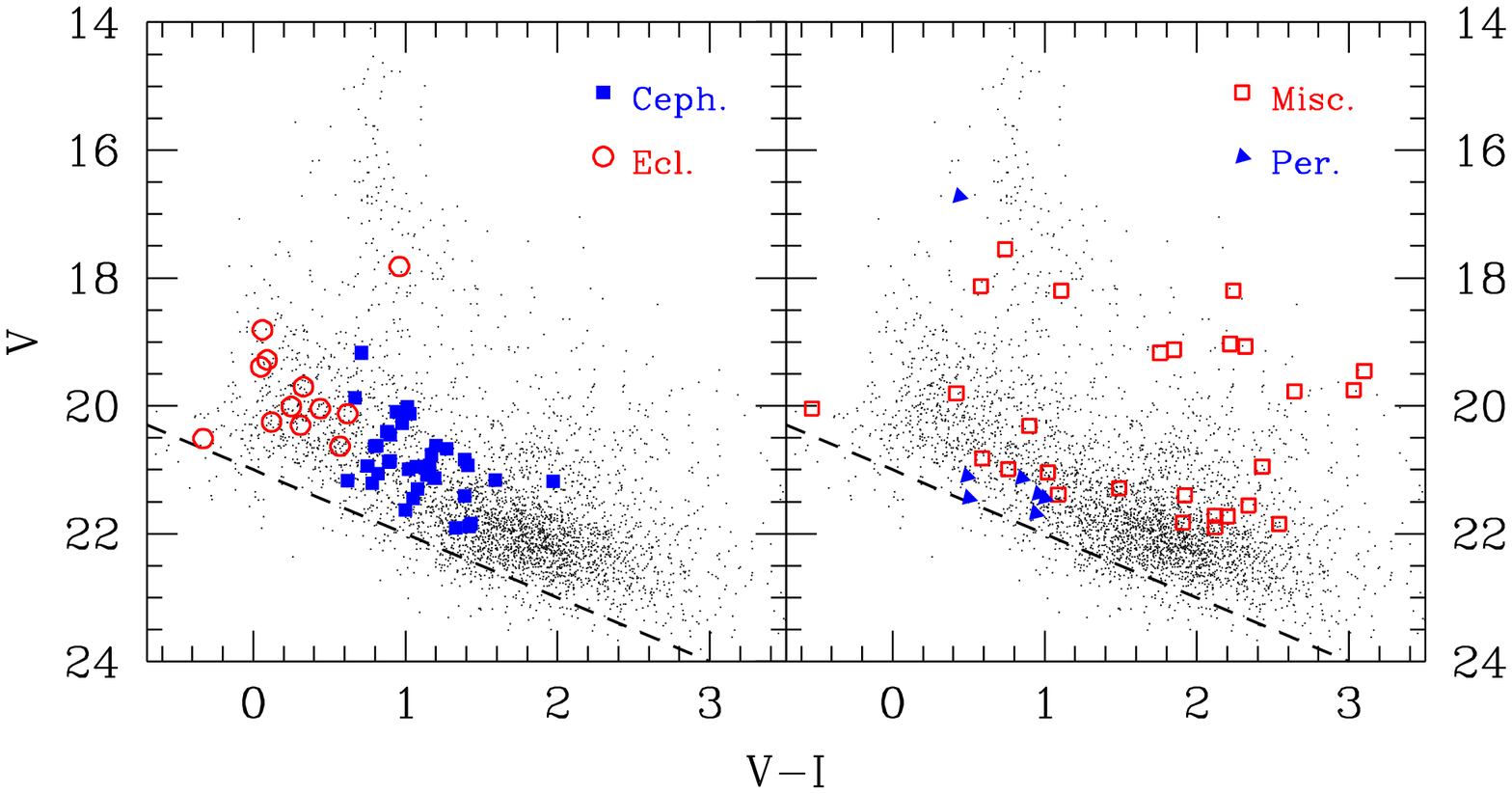}{8cm}{0}{85}{85}{-250}{-340}
\caption{$V,\;V-I$ color-magnitude diagrams for
the variable stars found in the field M31B. The eclipsing binaries and
Cepheids are plotted in the left panel and the other periodic
variables and miscellaneous variables are plotted in the right
panel. The dashed line corresponds to the $I$ detection limit of
$I\sim21\;{\rm mag}$.
\label{fig:cmd_var}}
\end{figure}

In Fig.\ref{fig:cmd_var} we show $V,\;V-I$ color-magnitude diagrams for
the variable stars found in the field M31B. The eclipsing binaries and
Cepheids are plotted in the left panel and the other periodic
variables and miscellaneous variables are plotted in the right panel.
As expected, the eclipsing binaries, with the exception of the
foreground W UMa system V438 D31B, occupy the blue upper main sequence
of M31 stars. Also as expected, the Cepheid variables group near
$V-I\sim1.0$, with the exception of possibly highly reddened system
V7713 D31B. The other periodic variable stars have positions on the
CMD similar to the Cepheids, again with the exception of the
foreground RR Lyr V7553 D31B.  The miscellaneous variables are
scattered throughout the CMD and clearly represent many classes of
variability, but most of them are red with $V-I=1.6-3.2$, and are
probably Mira variables.

\begin{figure}[t]
\plotfiddle{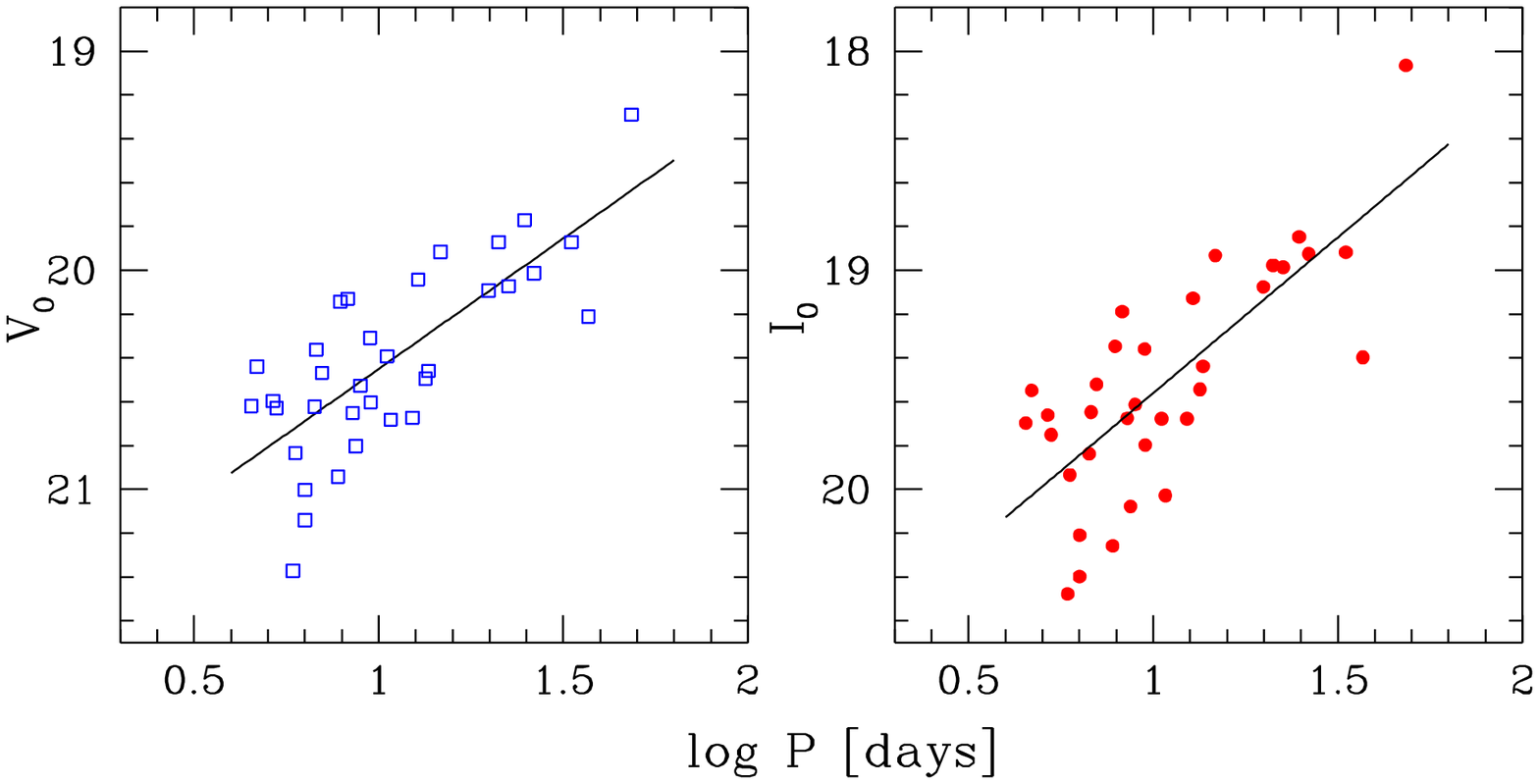}{8cm}{0}{85}{85}{-250}{-340}
\caption{The PL relations in the $V$ and $I$ bands for 34 Cepheids in 
field M31B. A preliminary estimate of the extinction has been used
(see text for more details).}
\label{fig:pl}
\end{figure}

The classical Cepheids found span pulsation periods from 4.5 to
$57\;days$ and all of them appear to be fundamental mode
pulsators. The period-luminosity [PL] relations (in the $V$ and $I$
bands) for the 34 Cepheids in field M31B are shown in
Fig.\ref{fig:pl}. They resemble the PL relations in Field III of
Freedman \& Madore (1990), which contain 16 Cepheids observed in
$BVRI$ filters. The distribution of the V and I residuals is also
similar to that found by Gould (1994) for Field III. Using the
technique described in Sasselov et al.~(1996), and adopting M31
foreground extinction $E(B-V)=0.08$ and no depth dispersion, we
estimate the mean extinction of the M31B Cepheid sample to be
$E(B-V)=0.2$. The range of individual extinctions is very large, up to
$E(B-V)=0.6$.  By enforcing positivity of the extinction, two of the
36 Cepheids are found to have luminous companions (or blends).  The
nominal distance difference between LMC and M31 from our sample is
$\Delta\mu(M31-LMC)=6.05\pm0.15$.  Due to the still small sample and
only two-band photometry these estimates are only preliminary; the
final sample from all fields should allow us to derive PL relations to
study dependencies as a function of galactocentric position and derive
the distance to M31.

At this stage of the DIRECT project we are interested mostly in
identifying interesting variable stars in M31 and M33.  As we
demonstrated 1-meter class telescopes are sufficient for this purpose,
providing one can obtain enough telescope time. During the next stage
of our project, the most promising detached eclipsing binaries and
Cepheid variables will be selected from our M31 and M33 variable star
catalogs to do accurate ($\sim1\%$) follow-up photometry in the $BVI$
bands.  A 2-meter class telescope with good seeing will be necessary
to obtain enough photometric precision. These accurate light curves
will then be used to determine the solutions of photometric orbits of
eclipsing binaries, a well-understood problem in astronomy (Wilson \&
Devinney 1971), as well for the modified Baade-Wesselink technique
modelling of the Cepheids (Krockenberger, Sasselov \& Noyes 1997).

Another step of this project, which requires obtaining high S/N radial
velocity curves to get the radii in physical units, would be realized
using one of the new 6.5-10 meter class telescopes.  For an idealized
DEB system of two identical mass stars viewed exactly in the orbital
plane, the expected semi-amplitude of the radial velocity curve is
given by
\begin{equation}
K=135\;\left(\frac{M_{star}}{M_{\odot}}\right)^{1/3}
       \left(\frac{P_{orbital}}{1\;day}\right)^{-1/3}\;km\;s^{-1},
\end{equation}
which for late O--early B type binaries typically translates to
$\sim200\;km\;s^{-1}$ (e.g. V478 Cyg or CW Cep, see Andersen 1991).
Determining the distance to an accuracy of 5\% requires knowing the
semi-amplitudes of the radial velocity curve to $\sim10\;km\;s^{-1}$
-- a very demanding, but not impossible task.

The last step would be the calculation of the distances: knowing the
surface brightness and the stellar radii of the DEB system or
Cepheid we can obtain the absolute stellar luminosities in the
observed band $F_{stellar~surface}$, and from the apparent fluxes
measured $F_{telescope}$ we can directly obtain the distance
\begin{equation}
d = \left( { F_{stellar~surface} \over F_{telescope} } \right) ^{1/2}
 R_{star} . 
\end{equation}
This means that we need very accurate absolute photometry from the
observed system in some selected band or, better, in several bands.
This also means that we have to be able to estimate the surface
brightness in some selected band of each star from the observed colors
or spectra.  Interstellar extinction is always a problem for any
photometric distance determination. To correct for that, multi-band
absolute photometry outside the eclipses in standard $UBVI$ and
possibly $JHK$ will be obtained. De-reddening for early type stars is
a standard and simple problem. As M33 is nearly a face-on system, the
problems with the interstellar extinction for this galaxy may be less
severe than for M31, a galaxy with obvious and patchy extinction.

\acknowledgments{We would like to thank the TAC of the 
Michigan-Dartmouth-MIT (MDM) Observatory and the TAC of the
F. L. Whipple Observatory (FLWO) for the generous amounts of telescope
time devoted to this project. We are very grateful to Bohdan
Paczy\'nski for motivating us to undertake this project and his always
helpful comments and suggestions. We thank Chris Kochanek for his very
careful reading and comments on the manuscript. Przemek Wo\'zniak
supplied us with FITS-manipulation programs we used to create the
finding charts. We thank the referee for very useful comments. The
staff of the MDM and the FLWO observatories is thanked for their
support during the long observing runs.  JK was supported by NSF grant
AST-9528096 to Bohdan Paczy\'nski and by the Polish KBN grant
2P03D011.12. KZS was supported by the Harvard-Smithsonian Center for
Astrophysics Fellowship.  JLT was supported by the NSF grant
AST-9401519.}

\end{document}